\documentclass[preprint,12pt,3p]{elsarticle}

\usepackage{amsmath,amssymb,amsfonts}
\usepackage{graphicx}
\usepackage{booktabs}
\usepackage{multirow}
\usepackage{xcolor}
\usepackage{hyperref}
\usepackage{lineno}
\usepackage{natbib}
\biboptions{sort&compress}
\usepackage{comment}

\journal{Chaos, Solitons \& Fractals}

\begin{document}
\begin{frontmatter}

\title{Multifractal Complexity of the Chandler Wobble and Its
Anomalous Disappearance (2015--2020): A MFDFA Study}

\author[1]{Sebasti\'an Jaroszewicz}
\author[2]{Nahuel Mendez\corref{cor1}}
\ead{nahueldanielmendez@gmail.com}
\author[3]{Maria P. Beccar-Varela}
\author[3]{Maria Cristina Mariani}

\address[1]{Comisi\'on Nacional de Energ\'ia At\'omica, Buenos Aires, Argentina}
\address[2]{Instituto S\'abato, Buenos Aires, Argentina}
\address[3]{Department of Mathematical Sciences, University of Texas at El Paso, El Paso, United States}

\cortext[cor1]{Corresponding author}

\begin{abstract}
The Chandler wobble (CW) --- the $\sim$433-day free nutation of
Earth's rotation pole --- experienced an anomalous near-disappearance
between 2015 and 2020, followed by a re-excitation with an
approximately $180^{\circ}$ phase reversal. Using Multifractal
Detrended Fluctuation Analysis (MFDFA) applied to more than
six decades (1962--2024) of daily IERS EOP C04 polar motion data,
this study provides the first multifractal characterisation of the CW
and its recent anomaly. Global MFDFA shows that the residual polar
motion components and the CW amplitude are genuine multifractal
processes with strongly $q$-dependent generalised Hurst exponents and
broad singularity spectra. Surrogate-data tests with shuffled and
phase-randomised ensembles demonstrate that this multifractality
originates from the combined action of long-range temporal
correlations and heavy-tailed excitation statistics. A sliding-window
analysis reveals a pronounced collapse in long-range persistence and
multifractal spectral width of the geometric polar motion signal
several years before and during the 2015--2020 amplitude minimum,
indicating a genuine dynamical regime change rather than a simple
suppression of oscillation amplitude. In contrast, the amplitude- and
phase-related variables retain broad multifractal spectra and stable
scaling exponents across all epochs, revealing a dynamical decoupling
between the geometry of the CW and the multiscale structure of its
amplitude and phase fluctuations. These findings highlight the CW
amplitude as an exceptionally multifractal integrator of geophysical
excitation and suggest that multifractal metrics may provide
early-warning indicators of major transitions in Earth rotation
dynamics.
\end{abstract}


\end{frontmatter}

%\linenumbers

%% ============================================================
%% ============================================================
\section{Introduction}
\label{sec:intro}

Earth's rotation axis does not remain fixed relative to the planet's
surface but traces a roughly circular path around the mean pole --- a
phenomenon known as polar motion. This motion is the superposition of
several components operating on distinct timescales: a secular drift
of the mean pole at $\sim$10~mas\,yr$^{-1}$ driven by glacial isostatic
adjustment and contemporary ice mass loss \cite{Adhikari2016}; a forced
annual oscillation excited by seasonal redistribution of atmospheric
and hydrological masses \cite{Lambeck}; and the Chandler wobble
(CW), a free nutation of the rotation pole about the figure axis with
a period of approximately 433~days and typical amplitudes of 100--200
milliarcseconds (mas) \cite{Gross2000}. Predicted theoretically by
Euler (1765) for a rigid Earth with a period of $\sim$305~days, the
wobble was first observed by Seth Carlo Chandler in 1891, with its
longer-than-predicted period later explained by the elastic yielding
of the Earth's mantle and the dynamical effects of the oceans and
atmosphere \cite{Gross2003}.

Despite over a century of study, the excitation mechanism sustaining
the Chandler wobble against viscous damping (characterised by a
quality factor $Q\sim50$--$100$, implying a damping time of
$\sim$30$\to$70~yr) remains an active subject of research \cite{Gross2000}.
Early proposals attributed the excitation to seismic activity,
large-scale atmospheric pressure variations, and geomagnetic jerks,
but quantitative estimates showed these sources to be insufficient
\cite{Aoyama2001}. A breakthrough came with Gross (2000), who
demonstrated that roughly two-thirds of the wobble excitation energy
originates from fluctuating pressure on the ocean floor caused by
wind-driven and thermohaline circulation, with the remaining
one-third attributable to atmospheric surface pressure \cite{Gross2000}.
This finding was later confirmed and refined by Gross et al.\ (2003)
using coupled ocean--atmosphere reanalyses \cite{Gross2003}. More
recently, hydrological contributions --- particularly variations in
continental water storage, ice sheets, and glaciers --- have been
recognised as non-negligible, especially on decadal timescales
\cite{Adhikari2016,Shi2025}.

The amplitude of the Chandler wobble is not constant but undergoes
pronounced modulation on interannual to decadal timescales, with
episodes of reduced amplitude separated by periods of vigorous
oscillation. The most dramatic such event in the instrumental record
occurred between approximately 2015 and 2020, when the CW amplitude
collapsed from its typical value of $\sim$150~mas to less than 10~mas
--- effectively making the wobble undetectable above the noise level
\cite{Zotov2023,Jeon2025}. This near-disappearance was followed by a
re-excitation around 2020--2021, remarkable for the approximately
$180^{\circ}$ reversal of the CW phase, an event previously observed
only in the 1920s--1930s \cite{Zotov2023}. The physical origin of the
2015--2020 quiescence has been debated: Zotov et al.\ (2024) proposed
a connection to anomalous changes in continental water mass
distribution and ocean circulation \cite{Zotov2024}; Shi et al.\ (2025)
and Shen et al.\ (2025) used GRACE/GRACE-FO satellite gravimetry to
attribute the quiescence primarily to hydrological and cryospheric
mass anomalies originating around 2011--2012, whose cumulative excitation
effect destructively interfered with the ongoing CW oscillation
\cite{Shi2025}. The subsequent phase reversal upon re-excitation
supports a picture in which the wobble amplitude was driven through
zero and re-emerged with inverted phase --- a behaviour consistent with
a driven damped oscillator subject to a change in the sign of the
forcing \cite{Zotov2024}.

From a dynamical systems perspective, the Earth's polar motion is a
high-dimensional, nonlinear, nonstationary process driven by a
multitude of geophysical forcing mechanisms operating across a broad
range of spatial and temporal scales. Characterising the statistical
self-similarity and long-range correlations of polar motion time series
is therefore of fundamental interest, both for understanding the
underlying geophysical processes and for improving predictions of Earth
orientation parameters, which are critical for satellite navigation,
geodesy, and deep-space communications \cite{Gross2015}. Detrended
Fluctuation Analysis (DFA), introduced by Peng et al.\ \cite{Peng1994},
has proven to be a powerful tool for quantifying long-range correlations
in nonstationary time series, and has been applied to geophysical series
ranging from temperature records to seismic catalogues. However, many
geophysical systems exhibit \emph{multifractal} rather than simple
monofractal scaling, meaning that the scaling exponent varies with the
order $q$ of the statistical moments considered. This richer structure
carries information about the intermittency and hierarchical organisation
of the underlying dynamics \cite{Kantelhardt2002}.

Multifractal Detrended Fluctuation Analysis (MFDFA), introduced by
Kantelhardt et al.\ (2002) \cite{Kantelhardt2002}, extends the DFA
framework by computing the $q$-th order fluctuation function $F_q(s)$
across a range of temporal scales $s$. The resulting generalised
Hurst exponent spectrum $h(q)$, mass exponent $\tau(q)$, and
singularity spectrum $f(\alpha)$ provide a complete statistical
characterisation of the multiscale structure of the analysed signal.
MFDFA has been applied to a wide range of physical, biological, and
economic time series \cite{Kantelhardt2002,Telesca2004}, but --- to
the best of our knowledge --- has never been applied to polar motion
or the Chandler wobble. The key question motivating this study is
whether the anomalous 2015--2020 quiescence of the CW is accompanied
by a change in the multifractal structure of polar motion, i.e.,
whether it represents not merely an amplitude suppression but a
genuine dynamical regime change detectable through the scaling
properties of the time series.

An additional methodological challenge concerns the origin of
observed multifractality. As shown by Kantelhardt et al.\ (2002) and
Schreiber and Schmitz (2000), multifractality in empirical time series
can arise from two distinct sources: long-range temporal correlations
(LRC) in the series, or a broad, heavy-tailed probability distribution
of the series values (fat tails), or a combination of both
\cite{Kantelhardt2002,Schreiber2000}. Distinguishing between these
sources is physically important --- LRC multifractality reflects
memory and correlation structure in the forcing process, while fat-tail
multifractality reflects the probability distribution of excitation
amplitudes. We address this question systematically using surrogate
data tests.

This paper presents the first MFDFA characterisation of the Chandler wobble and polar motion using more than six decades of daily IERS EOP
C04 measurements. The primary objectives of this study are twofold. First, we aim to establish the global multifractal properties of the CW residual signal and its amplitude and phase components, verifying through surrogate tests whether the observed multifractality is genuine and identifying its dynamical sources, such as long-range correlations or heavy-tailed probability distributions. Second, we seek to quantify the temporal evolution of multifractal complexity through a sliding-window MFDFA across five physically motivated temporal epochs spanning 1965-2024. This comprehensive approach allows us to test whether the anomalous 2015-2020 period is statistically distinguishable from prior epochs, and to critically assess whether these shifts in multiscale complexity provide potential early-warning signatures of the quiescence event.

The remainder of this paper is organised as follows. Section~2
describes the data and preprocessing pipeline. Section~3 presents the
MFDFA methodology, surrogate test protocol, and sliding-window
framework. Section~4 reports the results. Section~5 discusses the
physical interpretation. Section~6 summarises the conclusions.

%% ============================================================
\section{Data and Preprocessing}
\label{sec:data}

\subsection{IERS EOP C04 Series}

The primary data source is the IERS Earth Orientation Parameters C04
operational series (EOP 14 C04), maintained by the Paris Observatory
IERS centre and freely available at
\url{https://hpiers.obspm.fr/iers/eop/eopc04/} \cite{Adhikari2016}.
This series provides daily measurements of polar motion components
$x_p$ and $y_p$ (in arcseconds), Universal Time $\mathrm{UT1}-\mathrm{UTC}$
(in seconds), and length-of-day (LOD) anomalies, derived from a
combination of space-geodetic techniques including Very Long Baseline
Interferometry (VLBI), satellite and lunar laser ranging (SLR/LLR),
and Global Navigation Satellite Systems (GNSS) \cite{Gross2003}.
The time span used in this study is 1 January 1962 to 31 December 2024,
yielding a total of $N=22{,}831$ daily samples. The formal measurement
uncertainty in the polar motion components is of order
0.05--0.10~mas throughout most of the record, decreasing to
$<$0.05~mas after the introduction of GPS-based solutions in the
early 1990s.

\begin{figure}[htbp]
  \centering
  \includegraphics[width=\linewidth]{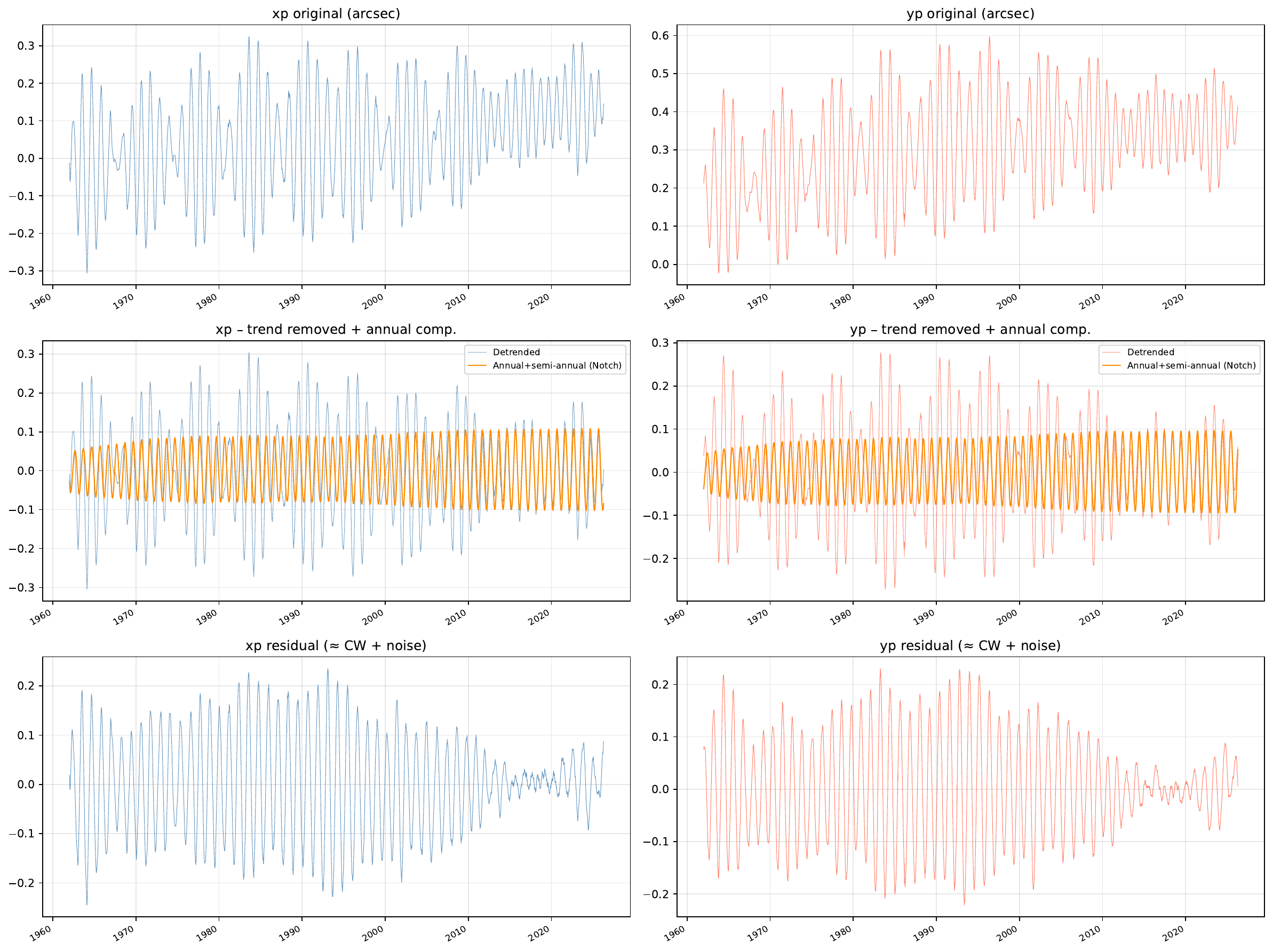}
  \caption{Step~1.2 --- Decomposition of polar motion from the
  IERS EOP C04 series (1962--2024). Top row: raw $x_p$ (left) and
  $y_p$ (right) components in arcseconds. Middle row: detrended
  series (blue) after removal of the secular mean-pole drift,
  together with the isolated annual+semi-annual component obtained
  by narrow-band notch filtering (orange), for $x_p$ (left) and
  $y_p$ (right). Bottom row: residual polar motion
  $x_p^{\rm res}$ (left) and $y_p^{\rm res}$ (right), which contain
  predominantly the Chandler wobble plus broadband stochastic noise
  and are the starting point for the multifractal analysis.}
  \label{fig:decomp}
\end{figure}

The raw $x_p$ and $y_p$ series are plotted in
Fig.~\ref{fig:decomp}, which illustrates the three dominant
components: the secular drift of the mean pole, the annual wobble,
and the Chandler wobble. The latter is clearly visible as the
dominant quasi-periodic oscillation with a period of $\sim$433~days,
whose amplitude undergoes pronounced decadal modulation including the
near-disappearance during 2015--2020.

\subsection{Removal of Secular Drift and Annual Signal}

Three deterministic components were systematically removed from the raw polar motion series prior to multifractal analysis, following standard geodetic practice\cite{Gross2015}. First, the secular mean-pole drift was modeled by a second-degree polynomial fitted by ordinary least squares to the full series and subsequently subtracted, leaving residuals with no significant linear or quadratic trends (Augmented Dickey-Fuller test, $p<0.01$). Next, the forced annual wobble, driven primarily by seasonal redistributions of atmospheric and hydrological masses \cite{Adhikari2016}, was eliminated using a zero-phase infinite-impulse-response (IIR) notch filter of order 4 centered at $f_A=1.000$ cpy with a quality factor $Q=30$. The zero-phase
implementation (forward--backward filtering) avoids any phase distortion in the residual signal.Finally, a smaller semi-annual component at $f_{SA}=2.000$, attributable to   elliptical terms in the atmospheric and oceanic forcing, cpy was removed using a second notch filter of identical order and quality factor. This zero-phase filtering approach avoids phase distortion, ensuring that the resulting residual series $x_p^{res}(t)$ and $y_p^{res}(t)$ retain the Chandler wobble as their dominant component alongside interannual variability and broadband stochastic noise.

\begin{figure}[htbp]
  \centering
  \includegraphics[width=0.85\linewidth]{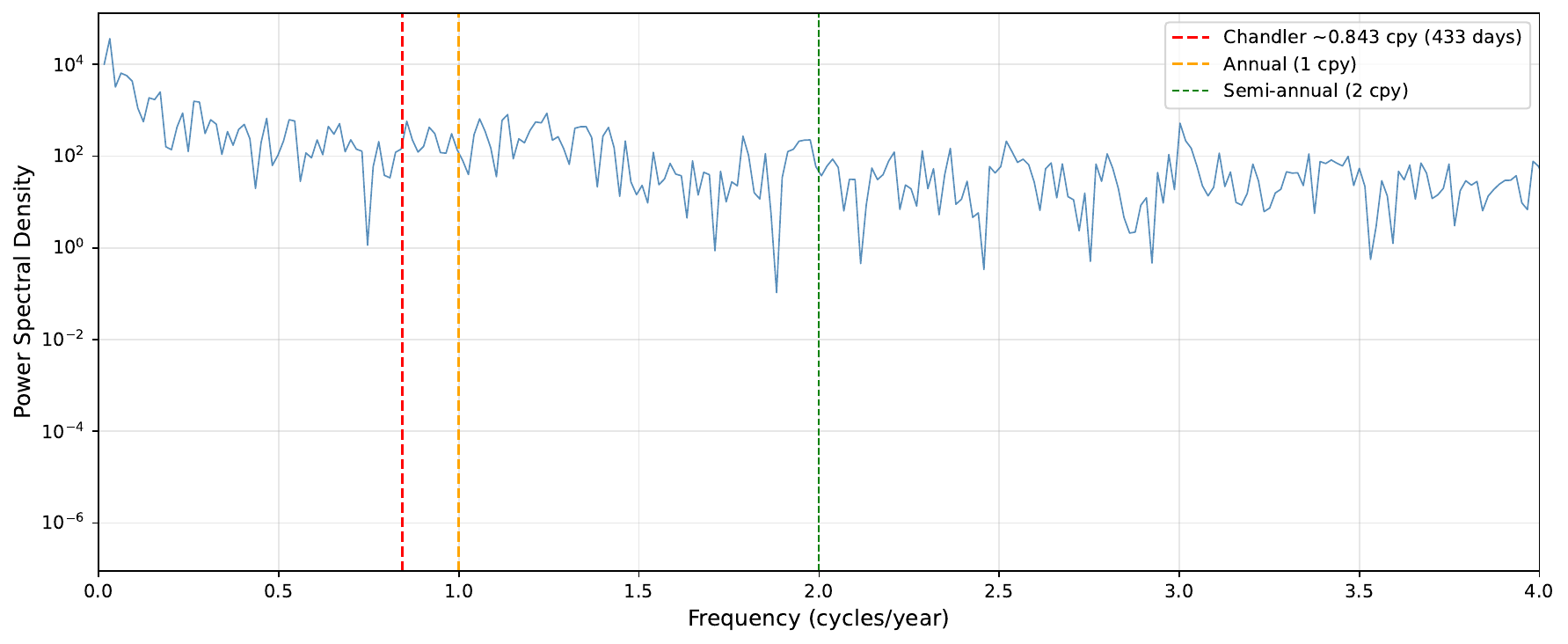}
  \caption{Power spectral density (PSD) of the complex residual
  polar motion $p^{\rm res}(t)=x_p^{\rm res}+iy_p^{\rm res}$
  after removal of secular drift and annual/semi-annual signals.
  The dominant peak at $f\approx0.843$~cpy ($T\approx433$~days)
  corresponds to the Chandler wobble. Vertical dashed lines mark
  the bandpass filter limits ($0.743$--$0.943$~cpy) used for
  CW extraction.}
  \label{fig:spectrum}
\end{figure}

The resulting residual series $x_p^{\rm res}(t)$ and $y_p^{\rm res}(t)$ retain the Chandler wobble as their dominant
component, together with interannual variability, broadband stochastic excitation, and measurement noise. The power spectral density of the
complex residual $p^{\rm res}(t)=x_p^{\rm res}(t)+iy_p^{\rm res}(t)$ is shown in Fig.~\ref{fig:spectrum}, where the sharp spectral peak at
$f\approx0.843$~cpy ($T\approx433$~days) unambiguously identifies the Chandler wobble frequency. The finite width of this peak reflects the
amplitude modulation of the CW on decadal timescales; the peak amplitude decreased from its 1990--2010 maximum to effectively zero
during 2015--2020 \cite{Zotov2023}.

\subsection{Chandler Wobble Extraction}
\label{sec:cw_extraction}

The Chandler wobble signal was isolated by applying a linear-phase
finite-impulse-response (FIR) bandpass filter to both $x_p^{\rm res}$
and $y_p^{\rm res}$. The filter was designed using a Hamming window
of length $L=1001$ taps, centred at $f_{\rm CW}=0.843$~cpy with a
half-power bandwidth of $\pm0.10$~cpy, covering the period range
approximately 395--475~days. This bandwidth is wide enough to capture
the observed small variations in the Chandler period
($\pm$1--5~days \cite{Gross2015}) while excluding the annual wobble
with a stopband attenuation $>$60~dB. The linear-phase property of
the FIR design ensures zero group delay, i.e.\ no phase distortion is
introduced in the filtered signal.

The filtered complex CW signal is defined as:
\begin{equation}
  p_{\rm CW}(t) = x_{p,{\rm CW}}(t) + i\,y_{p,{\rm CW}}(t),
  \label{eq:complex}
\end{equation}
where $x_{p,{\rm CW}}$ and $y_{p,{\rm CW}}$ are the bandpass-filtered
components. The instantaneous amplitude and phase were then obtained
via the Hilbert transform:
\begin{equation}
  A_{\rm CW}(t) = |p_{\rm CW}(t)|, \qquad
  \phi_{\rm CW}(t) = \arg\bigl[p_{\rm CW}(t)\bigr].
  \label{eq:hilbert}
\end{equation}

Figure~\ref{fig:amplitude} presents $A_{\rm CW}(t)$ and the unwrapped
phase $\phi_{\rm CW}(t)$ over the full record.

The extracted amplitude time series $A_{CW}(t)$ exhibits several noteworthy dynamical features over the observational record. Following a broad maximum between approximately 1990 and 2010 with peak amplitudes reaching $\sim$200~mas, the signal underwent a progressive decline from 2010 to 2015. This decline culminated in an anomalous near-zero episode from approximately 2016 to 2020, during which $A_{CW} < 20$~mas for an extended period, representing the most pronounced amplitude quiescence in the 62-year record. Subsequently, a re-excitation phase began around 2020-2021. Remarkably, the phase re-emerged at a value approximately $180^\circ$ different from the pre-quiescence phase, a behavior that is dynamically consistent with a driven damped oscillator having passed through a nodal point \cite{Zotov2023,Shi2025}.

These features are quantified in Table~\ref{tab:segments} (Section~3)
and form the physical basis for the temporal segmentation used in the
statistical analysis.

\begin{figure}[htbp]
  \centering
  \caption{Chandler wobble amplitude and phase extracted from the
  IERS EOP C04 series. Top panel: instantaneous amplitude
  $A_{\rm CW}(t)=|p_{\rm CW}(t)|$ (mas) obtained from the
  bandpass-filtered complex signal $p_{\rm CW}(t)$. Bottom panel:
  unwrapped phase $\phi_{\rm CW}(t)=\arg[p_{\rm CW}(t)]$ (radians).
  The grey band marks the 2015--2020 interval of near-complete
  amplitude collapse, during which $A_{\rm CW}(t)$ drops below
  $\sim$20~mas, followed by a re-excitation with an approximate
  $180^{\circ}$ phase reversal around 2020.}
  \label{fig:amplitude}
\end{figure}

\subsection{Series prepared for multifractal analysis}

Four time series derived from the preprocessing pipeline are
submitted to MFDFA:
\begin{enumerate}
  \item $x_p^{\rm res}(t)$: residual of the $x_p$ component after
  removal of secular drift, annual, and semi-annual signals.
  \item $y_p^{\rm res}(t)$: analogous residual for the $y_p$
  component.
  \item $\delta A_{\rm CW}(t) \equiv dA_{\rm CW}/dt$: daily
  increment of the CW amplitude envelope.
  \item $\phi_{\rm res}(t)$: phase residual after removal of the
  linear phase trend corresponding to uniform rotation at
  $f_{\rm CW}$.
\end{enumerate}

Table~\ref{tab:series} summarises basic statistics of these four
series.

\begin{table}[htbp]
\caption{Basic statistics of the four series submitted to MFDFA.
All series span 1962--2024. For $\delta A_{\rm CW}$, statistics are
computed after removing outliers with $|z|>5$ to avoid domination
by a few extreme jumps at filter edges.}
\label{tab:series}
\centering\small
\begin{tabular}{lcccc}
\toprule
Series & Mean & Std & Skewness & Excess kurtosis\\
\midrule
$x_p^{\rm res}$ (arcsec) &
$2.62\times10^{-4}$ & $9.92\times10^{-2}$ & $-0.01$ & $-0.72$\\
$y_p^{\rm res}$ (arcsec) &
$1.32\times10^{-4}$ & $9.79\times10^{-2}$ & $0.07$ & $-0.74$\\
$\delta A_{\rm CW}$ (mas\,d$^{-1}$) &
$0.0$              & $4.61\times10^{-2}$ & $8.05$ & $2.77\times10^{2}$\\
$\phi_{\rm res}$ (rad) &
$3.26\times10^{-12}$ & $2.82\times10^{-1}$ & $0.53$ & $0.75$\\
\bottomrule
\end{tabular}
\end{table}

In addition, we define five temporal segments for the statistical
analysis, based on the known evolution of the Chandler wobble
amplitude (Fig.~\ref{fig:amplitude}). These segments are used later
in the sliding-window and hypothesis-testing procedures
(Section~\ref{sec:stats_method}) and are summarised in
Table~\ref{tab:segments}.

\begin{table}[htbp]
\caption{Temporal segments for statistical analysis. The number of
windows $n$ refers to the sliding-window MFDFA with window
$W=6$~yr and step $\Delta t=0.5$~yr (Section~\ref{sec:sliding_method}).}
\label{tab:segments}
\centering\small
\begin{tabular}{llrl}
\toprule
Segment & Period & $n$ & Physical characterisation\\
\midrule
Pre-active     & 1965--1990 & 50 & Moderate CW amplitude, variable\\
Active         & 1990--2010 & 40 & High sustained amplitude (150--200~mas)\\
Decline        & 2010--2015 & 10 & Systematic amplitude decrease\\
Anomalous      & 2015--2020 & 11 & Near-complete disappearance ($<$20~mas)\\
Post-anomalous & 2020--2024 &  5 & Re-excitation with $180^{\circ}$ phase reversal\\
\bottomrule
\end{tabular}
\end{table}

%% ============================================================
\section{Methods}
\label{sec:methods}

\subsection{Overview of the analysis pipeline}

The proposed analytical pipeline is structured sequentially. It begins with the preprocessing of the IERS EOP C04 time series to extract the complex Chandler wobble signal $p_{CW}(t)$, along with its instantaneous amplitude $A_{CW}(t)$ and phase $\phi_{CW}(t)$. From these primary components, four derived series ($x_p^{res}$, $y_p^{res}$, $\delta A_{CW}$, $\phi_{res}$) are constructed to serve as the foundation for the subsequent multifractal analysis. We then perform a global MFDFA on each series to robustly estimate the generalized Hurst exponents $h(q)$, mass exponents $\tau(q)$, and singularity spectra $f(\alpha)$ over the full 1962-2024 record. To rigorously validate our findings, surrogate-data tests are systematically applied to determine whether the observed multifractality is genuine and to distinguish between long-range correlations and fat-tailed probability distributions as its primary dynamical sources. Finally, a sliding-window MFDFA framework is applied to characterize the non-stationary temporal evolution of the multifractal complexity and to compute summary statistics across five physically motivated temporal segments.

All numerical computations were performed in Python (NumPy, SciPy, pandas) with double-precision arithmetic. Linear regressions were
carried out in log--log space using ordinary least squares. Only fits with coefficient of determination $R^2>0.98$ were retained
for interpretation.

\subsection{Multifractal Detrended Fluctuation Analysis}
\label{sec:mfdfa_method}

For a time series $\{x_k\}_{k=1}^N$, MFDFA proceeds as follows
\cite{Kantelhardt2002,Kantelhardt2001}:

\begin{enumerate}
  \item \textit{Profile construction.}
  Compute the integrated profile
  \begin{equation}
    Y(i) = \sum_{k=1}^{i}\bigl[x_k - \langle x\rangle\bigr],
    \qquad i=1,\dots,N,
    \label{eq:profile}
  \end{equation}
  where $\langle x\rangle$ is the sample mean. This step maps
  stationary long-range correlated series to a nonstationary random
  walk whose fluctuations can be more robustly characterised.

  \item \textit{Segmentation.}
  For a given scale $s$ (integer number of samples), divide $Y(i)$
  into $N_s=\lfloor N/s\rfloor$ non-overlapping segments of length
  $s$ starting from the beginning, and another $N_s$ segments
  starting from the end (backward segmentation), yielding a total
  of $2N_s$ segments. This symmetric segmentation improves
  statistical robustness for large $s$.

  \item \textit{Local detrending.}
  In each segment $\nu=1,\dots,2N_s$, fit a polynomial trend of order
  $m$ and compute the variance of the detrended profile:
  \begin{equation}
    F^2(\nu,s) = \frac{1}{s}
    \sum_{i=1}^{s}\left\{
      Y\bigl[(\nu-1)s+i\bigr] - y_{\nu,m}(i)
    \right\}^2,
    \label{eq:variance}
  \end{equation}
  where $y_{\nu,m}(i)$ is the least-squares polynomial of degree
  $m$ fitted to the profile in segment $\nu$. In this work we use
  $m=2$ (MFDFA-2), which efficiently removes linear and quadratic
  trends while avoiding overfitting small-scale fluctuations.

  \item \textit{$q$-order fluctuation function.}
  For a real parameter $q$, define the $q$-th order fluctuation
  function
  \begin{equation}
    F_q(s)=\left\{
      \frac{1}{2N_s}\sum_{\nu=1}^{2N_s}
        \bigl[F^2(\nu,s)\bigr]^{q/2}
    \right\}^{1/q},
    \qquad q\neq0,
    \label{eq:Fq_again}
  \end{equation}
  with the logarithmic definition for $q=0$:
  \begin{equation}
    F_0(s) = \exp\left\{
      \frac{1}{4N_s}\sum_{\nu=1}^{2N_s}
      \ln\bigl[F^2(\nu,s)\bigr]
    \right\}.
  \end{equation}
  Negative values of $q$ emphasise small fluctuations, whereas
  positive values emphasise large fluctuations.

  \item \textit{Scaling law and generalised Hurst exponent.}
  For a multifractal process, the fluctuation functions obey a
  power-law scaling
  \begin{equation}
    F_q(s)\sim s^{h(q)},
    \label{eq:scaling}
  \end{equation}
  where $h(q)$ is the generalised Hurst exponent. The exponent is
  estimated as the slope of $\log_{2} F_q(s)$ vs\ $\log_{2} s$
  obtained by linear regression over a range of scales
  $s\in[s_{\min},s_{\max}]$.

  \item \textit{Mass exponent and singularity spectrum.}
  The mass exponent spectrum is defined as
  \begin{equation}
    \tau(q)=q\,h(q)-1,
  \end{equation}
  and the singularity (multifractal) spectrum $f(\alpha)$ is
  obtained via Legendre transform:
  \begin{equation}
    \alpha=\frac{d\tau}{dq}, \qquad
    f(\alpha)=q\,\alpha - \tau(q).
  \end{equation}
  Practically, $\alpha(q)$ is obtained by discrete differentiation
  of $\tau(q)$, and $f(\alpha)$ is then computed for each $q$.
  The width of the singularity spectrum,
  $\Delta\alpha=\alpha_{\max}-\alpha_{\min}$, quantifies the degree
  of multifractality.
\end{enumerate}

In all analyses we use the scale range
$s\in[20,\,N/4]$ for the global MFDFA, with $N$ the length of the
series under consideration. Scales are sampled logarithmically using
20--30 values equally spaced in $\log_{10}s$. The set of $q$-values
is chosen as
$q\in\{-8,-7,\dots,-1,0,1,\dots,8\}$, which provides adequate
coverage of both small- and large-fluctuation regimes without
amplifying numerical instability at very large $|q|$.

Because MFDFA is always applied to the integrated profile $Y(i)$,
the exponents $h(q)$ reported in this work refer to that profile.
For stationary fractional Gaussian noise (fGn) processes  this
implies the relation
$H_{\rm real} = h(q{=}2) - 1$ for the Hurst exponent of the original
series \cite{Kantelhardt2002}. In the present context, however,
polar motion residuals and their sliding-window subsets are only
approximately stationary and contain a strong quasi-periodic CW
component. For this reason, we report and compare $H(t)\equiv h(q{=}2,t)$
directly as a local persistence metric, and use the global
$H_{\rm real}=h(2)-1$ values only as a rough reference. We do not
re-interpret sliding-window exponents as exact Hurst exponents of a
stationary process.

All multifractal analyses in this work were performed using the
\texttt{MF-toolkit} high-performance Python library for Multifractal
Detrended Fluctuation Analysis (MFDFA), which provides automated
crossover detection and surrogate-based source identification of
multifractality \cite{Mendez2026_MFtoolkit}. This codebase has been
previously validated in diverse applications, including the
multifractal analysis of historical humpback whale song recordings
\cite{Mendez2026_humpback} and a finite-size scaling protocol for MFDFA
in the 2D Ising model \cite{Jaroszewicz2026_IsingFSS}, ensuring the
robustness and reproducibility of the present results.

\subsection{Surrogate-data tests}
\label{sec:surrogates_method}

Observed multifractality in empirical time series may arise from two distinct sources: (i)~long-range temporal correlations (LRC) in
an otherwise approximately Gaussian process, and (ii)~a broad, heavy-tailed probability distribution (fat tails) in the absence of
non-trivial correlations \cite{Kantelhardt2002,Schreiber2000}.

To distinguish between these sources and rigorously test whether the observed multifractality in the polar motion series is genuine rather than a finite-sample artifact, we employ two distinct classes of surrogate data \cite{Schreiber2000}. First, randomly shuffled surrogates (RS) are generated by randomly permuting the original series, a procedure that explicitly destroys all temporal correlations while exactly preserving the empirical amplitude distribution. Consequently, any residual multifractality observed in these shuffled ensembles must originate entirely from heavy-tailed distributions. Second, we generate phase-randomized surrogates using the iterative amplitude-adjusted Fourier transform (IAAFT) algorithm. This method randomizes the phases of the discrete Fourier transform while preserving the power spectrum, and iteratively adjusts the amplitudes to match the original distribution. In the limit of convergence, this technique preserves the linear correlation structure but tends to Gaussianize the distribution of increments, ensuring that any residual multifractality strictly reflects long-range correlation (LRC) effects.

For each series analysed ($x_p^{\rm res}$, $y_p^{\rm res}$, $A_{\rm CW}$), we generate $N_{\rm surr}=20$ realisations of each
surrogate type. The full MFDFA pipeline is run on each surrogate, using the same set of scales and $q$-values as for the original
series. The multifractal spectral width $\Delta\alpha$ is computed for each surrogate, and its distribution is summarised by the mean
$\langle\Delta\alpha_{\rm surr}\rangle$ and standard deviation $\sigma(\Delta\alpha_{\rm surr})$.

Significance is quantified via the $Z$-score
\begin{equation}
  Z = \frac{\Delta\alpha_{\rm orig} -
  \langle\Delta\alpha_{\rm surr}\rangle}
  {\sigma(\Delta\alpha_{\rm surr})}.
  \label{eq:zscore_method}
\end{equation}
We interpret $Z>2$ as statistically significant at approximately the 95\,\% confidence level. If both $Z_{\rm sh}$ and $Z_{\rm pr}$ exceed
this threshold, the multifractality is classified as having a \emph{mixed} origin (LRC + fat tails).

\begin{figure}[htbp]
  \centering
  \includegraphics[width=\linewidth]{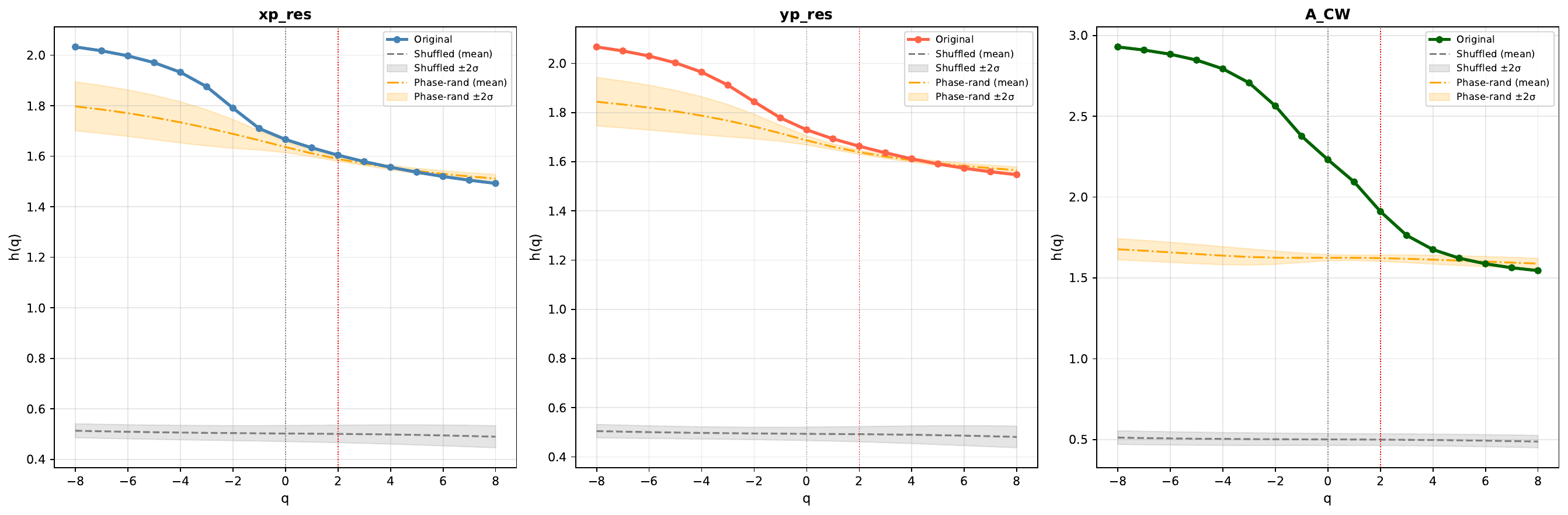}
  \caption{Generalised Hurst exponent $h(q)$ for the original series
(solid lines) and the mean $\pm$ one standard deviation of the
shuffled (RS, dashed) and phase-randomised (IAAFT, dotted)
surrogate ensembles ($N_{\rm surr}=20$ per type) for
$x_p^{\rm res}$, $y_p^{\rm res}$ and $A_{\rm CW}$. Shuffling
strongly reduces the $q$-dependence of $h(q)$, while
phase-randomisation produces intermediate behaviour, confirming
that both long-range correlations and heavy-tailed amplitude
distributions contribute to the observed multifractality.}
  \label{fig:surrogate_hq}
\end{figure}

 Fig.~\ref{fig:surrogate_hq} compares the values of $h(q)$ for the surrogate ensembles and the original series, while Figure~\ref{fig:surrogate_falpha} (Section~\ref{sec:results}) shows the original $f(\alpha)$ curves together with the surrogate distributions.

\subsection{Sliding-window MFDFA}
\label{sec:sliding_method}

To characterise the temporal evolution of the multifractal properties of polar motion, we apply MFDFA in a sliding-window fashion to the
four derived series. Let $x(t_i)$ denote a given series sampled daily at times $t_i$. We consider overlapping windows of fixed length
$W=6$~yr ($\approx$2190~days), advanced in steps of $\Delta t=0.5$~yr ($\approx$183~days). For each window $w$, defined by the index range
$i\in[i_{\rm start}^{(w)},i_{\rm end}^{(w)}]$, we compute the full MFDFA and extract three scalar metrics:

\begin{enumerate}
  \item The local generalised Hurst exponent
  $H(t_w)=h(q{=}2,t_w)$, where $t_w$ is the central time of the
  window. This quantity measures the degree of long-range
  persistence at time $t_w$.

  \item The local multifractal spectral width
  $\Delta\alpha(t_w)=\alpha_{\max}(t_w)-\alpha_{\min}(t_w)$,
  quantifying the intensity of multifractality in that window.

  \item The local spectral asymmetry
  $B(t_w)=[\alpha_0(t_w)-\alpha_{\min}(t_w)]/\Delta\alpha(t_w)$,
  where $\alpha_0$ is the value at which $f(\alpha)$ attains its
  maximum. Values $B<0.5$ indicate spectra dominated by large
  fluctuations, whereas $B>0.5$ indicate dominance of small
  fluctuations.
\end{enumerate}

Windows in which fewer than four valid scales contribute to the
$h(q)$ estimation (e.g.\ due to numerical issues at very large $|q|$)
are discarded. This affects only the early part of the record where
data are sparse; the impact on the 1965--2024 interval of interest is
negligible. The resulting time series $H(t)$, $\Delta\alpha(t)$ and
$B(t)$ are then segmented according to Table~\ref{tab:segments}, and
their means, standard deviations, medians and interquartile ranges
are computed for each segment and compared via the Mann--Whitney
$U$ test (Section~\ref{sec:results})

The choice $W=6$~yr ($\approx$\,2190 daily samples) reflects a
deliberate compromise between two competing requirements. On the one
hand, resolving the temporal non-stationarity of the CW demands
windows short enough to track changes on interannual timescales; a
window of 6~yr spans roughly five complete Chandler cycles
($T_{\rm CW}\approx433$~d), which is sufficient to resolve the
dominant quasi-periodic modulation while remaining sensitive to
transitions such as the onset of the 2015--2020 amplitude collapse.
On the other hand, MFDFA requires a minimum number of data points to
populate the fluctuation functions $F_q(s)$ across the scale range
$s\in[s_{\min},s_{\max}]=[20,N_w/4]$ without severe finite-size
bias. With $N_w\approx2190$ samples, the usable scale range spans
approximately 1.3 decades (from 20 to $\sim$550 days), which
previous studies have shown to be adequate for reliable estimation
of $h(q)$ when $R^2>0.98$ is enforced \cite{Kantelhardt2002,Peng1994}.
To verify the robustness of our results with respect to the window
length, we repeated the analysis with $W=4$~yr and $W=8$~yr; the
qualitative features of the $H(t)$ and $\Delta\alpha(t)$ time series
--- in particular the pronounced decline during 2005--2020 in
$x_p^{\rm res}$ and $y_p^{\rm res}$ --- are preserved in both cases,
confirming that $W=6$~yr is not a critical parameter. Shorter windows
($W<4$~yr, fewer than 3 CW cycles) did produce noisier estimates and
occasional $R^2<0.98$ failures, and were therefore discarded.

Figures~\ref{fig:H_sliding} and \ref{fig:Da_sliding} display the
time evolution of $H(t)$ and $\Delta\alpha(t)$ respectively for all
four series, with the anomalous period 2015--2020 highlighted.

\subsection{Statistical comparisons}
\label{sec:stats_method}

The five temporal segments used in the statistical analysis are
summarised in Table~\ref{tab:segments}. For each series and each
sliding-window metric
$M\in\{H(t),\Delta\alpha(t),B(t)\}$ the window-centre times $t_w$
were assigned to these segments and the corresponding samples were
used to compute the mean, standard deviation, median and interquartile
range reported in Section~\ref{sec:results}.

Between-segment differences were assessed with the two-tailed
Mann--Whitney $U$ test, a non-parametric alternative to the
two-sample $t$-test that does not assume normality of the
distributions. Significance levels are encoded using the standard
notation $^*p<0.05$, $^{**}p<0.01$, $^{***}p<0.001$, while
``n.s.'' denotes non-significant results. Comparisons are performed
(i) between each segment and the anomalous period (2015--2020), and
(ii) between each segment and the active period (1990--2010),
which serves as a reference epoch of sustained large CW amplitude.

%% ============================================================
\section{Results}
\label{sec:results}

\subsection{Global multifractal properties}
\label{sec:results_global}

\begin{figure}[htbp]
  \centering
  \includegraphics[width=0.85\linewidth]{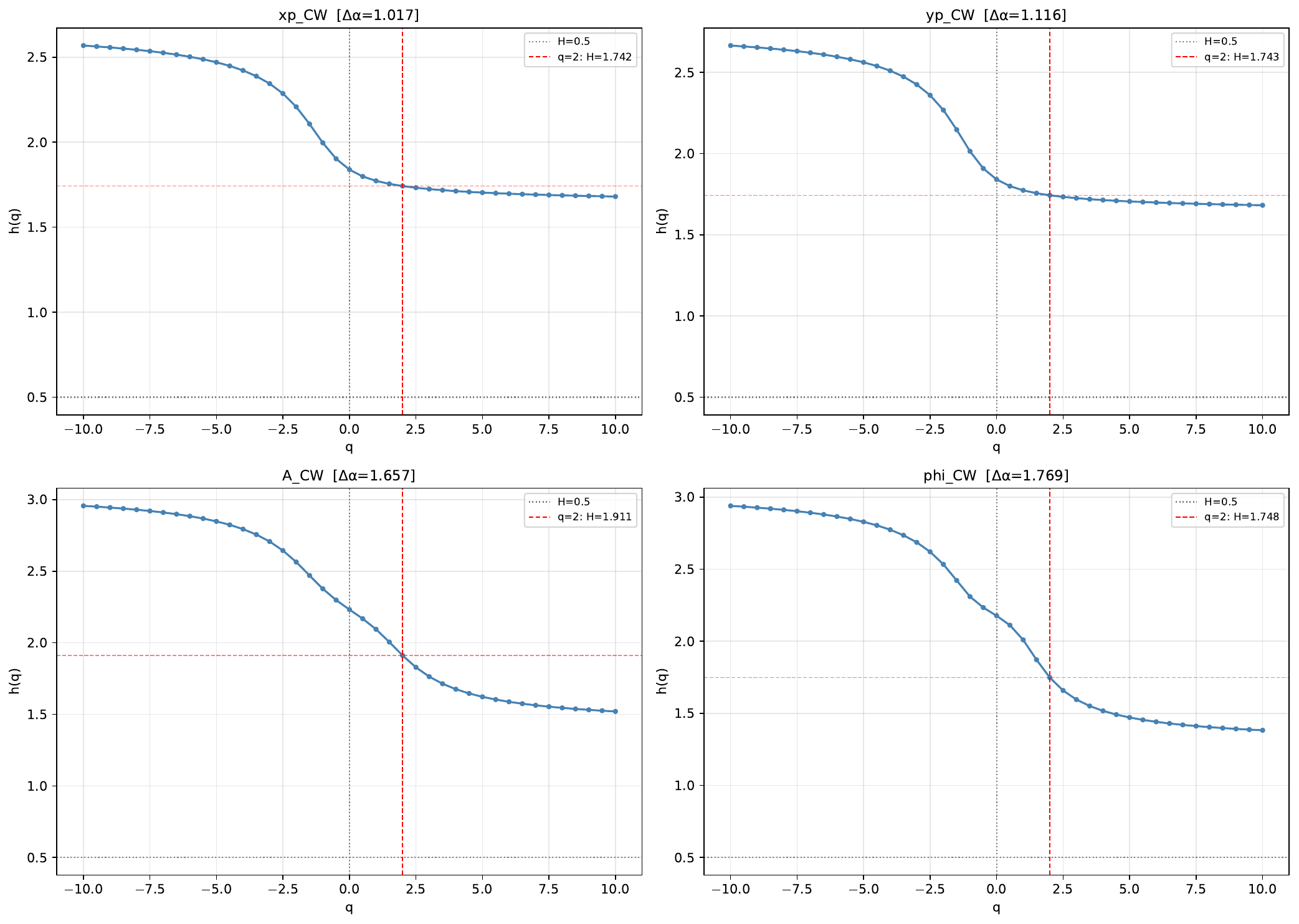}
  \caption{Generalised Hurst exponent $h(q)$ as a function of $q$
  for the four analysed series computed over the full 1962--2024
  record. The clear non-linearity of $h(q)$ and its strong
  $q$-dependence, especially for the amplitude series
  $A_{\rm CW}$ and $\delta A_{\rm CW}$, confirm the multifractal
  nature of the signals.}
  \label{fig:hq_global}
\end{figure}

Figure~\ref{fig:hq_global} displays the generalised Hurst exponent
$h(q)$ as a function of $q$ for the four analysed series. All curves
are markedly non-linear, with $h(q)$ decreasing as $q$ increases,
demonstrating the presence of multifractality. The amplitude-related
series, in particular $A_{\rm CW}$, exhibit a stronger $q$-dependence
than the geometric residuals, consistent with their broader
singularity spectra.

\begin{figure}[htbp]
  \centering
  \includegraphics[width=0.85\linewidth]{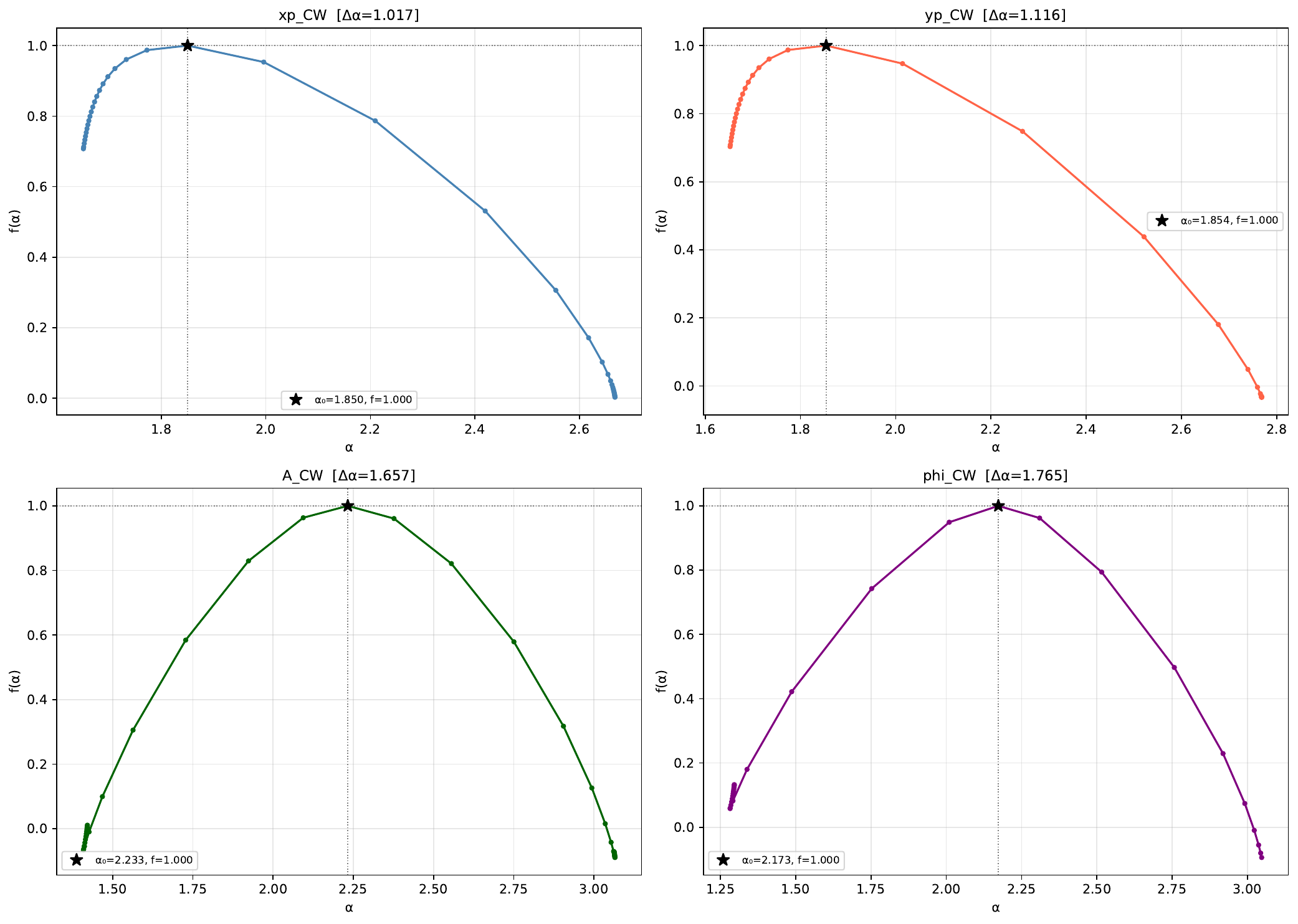}
  \caption{Global singularity spectra $f(\alpha)$ for the four
  analysed series computed from the full 1962--2024 record.
  The broad, non-parabolic shapes confirm genuine multifractality.
  The Chandler wobble amplitude $A_{\rm CW}$ displays the largest
  spectral width $\Delta\alpha$, indicating the strongest degree
  of multifractality.}
  \label{fig:falpha_global}
\end{figure}

Figure~\ref{fig:falpha_global} shows the singularity spectra
$f(\alpha)$ obtained from the global MFDFA of the four analysed
series ($x_p^{\rm res}$, $y_p^{\rm res}$, $A_{\rm CW}$,
$\delta A_{\rm CW}$) over the full 1962--2024 record. All spectra
are clearly non-parabolic and display finite width, indicating
genuine multifractality rather than simple monofractal scaling.
The amplitude envelope $A_{\rm CW}$ exhibits the broadest spectrum,
reflecting the richest multifractal structure.

Table~\ref{tab:global_mfdfa} summarises the main global MFDFA
parameters. The generalised Hurst exponent at $q=2$,
$h(q{=}2)\approx1.74$ for $x_p^{\rm res}$ and $y_p^{\rm res}$ and
$h(q{=}2)\approx1.91$ for $A_{\rm CW}$, corresponds to original-series
Hurst exponents $H_{\rm real}=h(2)-1$ in the range 0.74--0.91,
indicating strong long-range persistence. The multifractal spectral
widths are $\Delta\alpha\approx0.72$ for the polar motion residuals
and $\Delta\alpha\approx1.61$ for $A_{\rm CW}$, confirming that the
amplitude dynamics is considerably more complex than the geometric
polar motion.

\begin{table}[htbp]
\caption{Global MFDFA parameters for the four analysed series over
the full 1962--2024 record. $H_{\rm real}=h(q{=}2)-1$ denotes the
Hurst exponent of the original (non-integrated) series.}
\label{tab:global_mfdfa}
\centering\small
\begin{tabular}{lcccc}
\toprule
Series & $h(q{=}2)$ & $H_{\rm real}$ & $\Delta\alpha$ &
$\langle R^2\rangle$\\
\midrule
$x_p^{\rm res}$     & 1.742 & 0.742 & 0.737 & $>$0.99\\
$y_p^{\rm res}$     & 1.743 & 0.743 & 0.716 & $>$0.99\\
$A_{\rm CW}$        & 1.911 & 0.911 & 1.608 & $>$0.99\\
$\delta A_{\rm CW}$ & 1.590 & 0.590 & 1.030 & $>$0.99\\
\bottomrule
\end{tabular}
\end{table}

\subsection{Surrogate test results}
\label{sec:results_surrogates}

\begin{figure}[htbp]
  \centering
  \includegraphics[width=\linewidth]{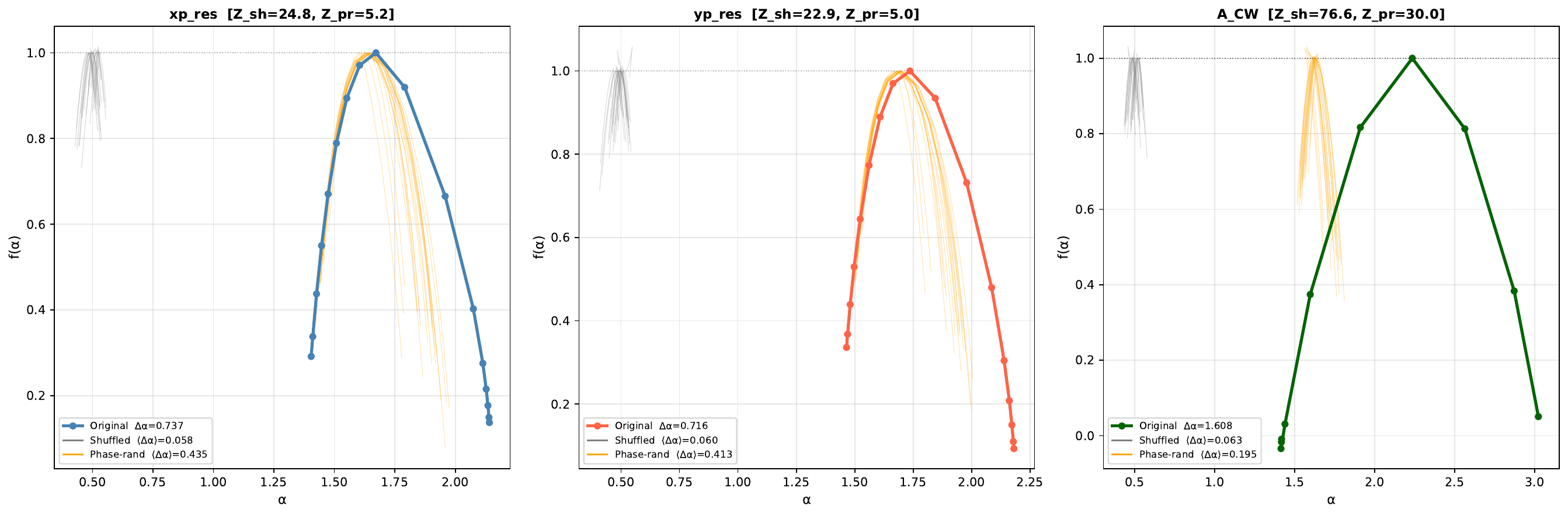}
  \caption{Singularity spectra $f(\alpha)$ for the original series
  (solid lines) and the mean $\pm$ one standard deviation of the
  shuffled (RS, dashed) and phase-randomised (IAAFT, dotted)
  surrogate ensembles ($N_{\rm surr}=20$ per type) for
  $x_p^{\rm res}$, $y_p^{\rm res}$ and $A_{\rm CW}$. The pronounced
  narrowing of $f(\alpha)$ in both surrogate classes confirms that
  the original multifractality is genuine.}
  \label{fig:surrogate_falpha}
\end{figure}

The surrogate-data analysis confirms that the multifractality of the
polar motion series is genuine and of mixed origin. Figure~\ref{fig:surrogate_falpha}
compares the original $f(\alpha)$ spectra with the mean and scatter
of the shuffled (RS) and phase-randomised (IAAFT) surrogates for
$x_p^{\rm res}$, $y_p^{\rm res}$ and $A_{\rm CW}$. Shuffling
dramatically narrows the spectra, while phase-randomisation produces
intermediate widths, demonstrating that both long-range correlations
and heavy-tailed amplitude distributions contribute to the observed
multifractality.

Table~\ref{tab:surrogate_results} lists the spectral widths for the
original and surrogate series and the corresponding $Z$-scores.
All three original series have $\Delta\alpha$ values far in excess of
those of the shuffled and phase-randomised surrogates. In particular,
$A_{\rm CW}$ exhibits extremely large $Z$-scores
($Z_{\rm sh}=76.6$, $Z_{\rm pr}=30.0$), underscoring the exceptional
multifractal richness of the CW amplitude dynamics.

\begin{table}[htbp]
\caption{Surrogate test results for $x_p^{\rm res}$,
$y_p^{\rm res}$ and $A_{\rm CW}$.
$\Delta\alpha_{\rm orig}$ is the spectral width of the original
series; $\langle\Delta\alpha_{\rm sh}\rangle$ and
$\langle\Delta\alpha_{\rm pr}\rangle$ are the mean surrogate widths
for shuffled (RS) and phase-randomised (IAAFT) ensembles
($N_{\rm surr}=20$ per type); $Z_{\rm sh}$ and $Z_{\rm pr}$ are the
corresponding $Z$-scores.}
\label{tab:surrogate_results}
\centering\small
\begin{tabular}{lcccccl}
\toprule
Series &
$\Delta\alpha_{\rm orig}$ &
$\langle\Delta\alpha_{\rm sh}\rangle\pm\sigma$ &
$Z_{\rm sh}$ &
$\langle\Delta\alpha_{\rm pr}\rangle\pm\sigma$ &
$Z_{\rm pr}$ &
Source of MF\\
\midrule
$x_p^{\rm res}$ &
  0.737 & $0.058\pm0.028$ & 24.8 &
  $0.435\pm0.076$ & 5.2 & LRC + fat tails\\
$y_p^{\rm res}$ &
  0.716 & $0.060\pm0.028$ & 22.9 &
  $0.413\pm0.079$ & 5.0 & LRC + fat tails\\
$A_{\rm CW}$ &
  1.608 & $0.063\pm0.020$ & 76.6 &
  $0.195\pm0.047$ & 30.0 & LRC + fat tails\\
\bottomrule
\end{tabular}
\end{table}

\subsection{Temporal evolution of $H(t)$}
\label{sec:results_H}

\begin{figure}[htbp]
  \centering
  \includegraphics[width=\linewidth]{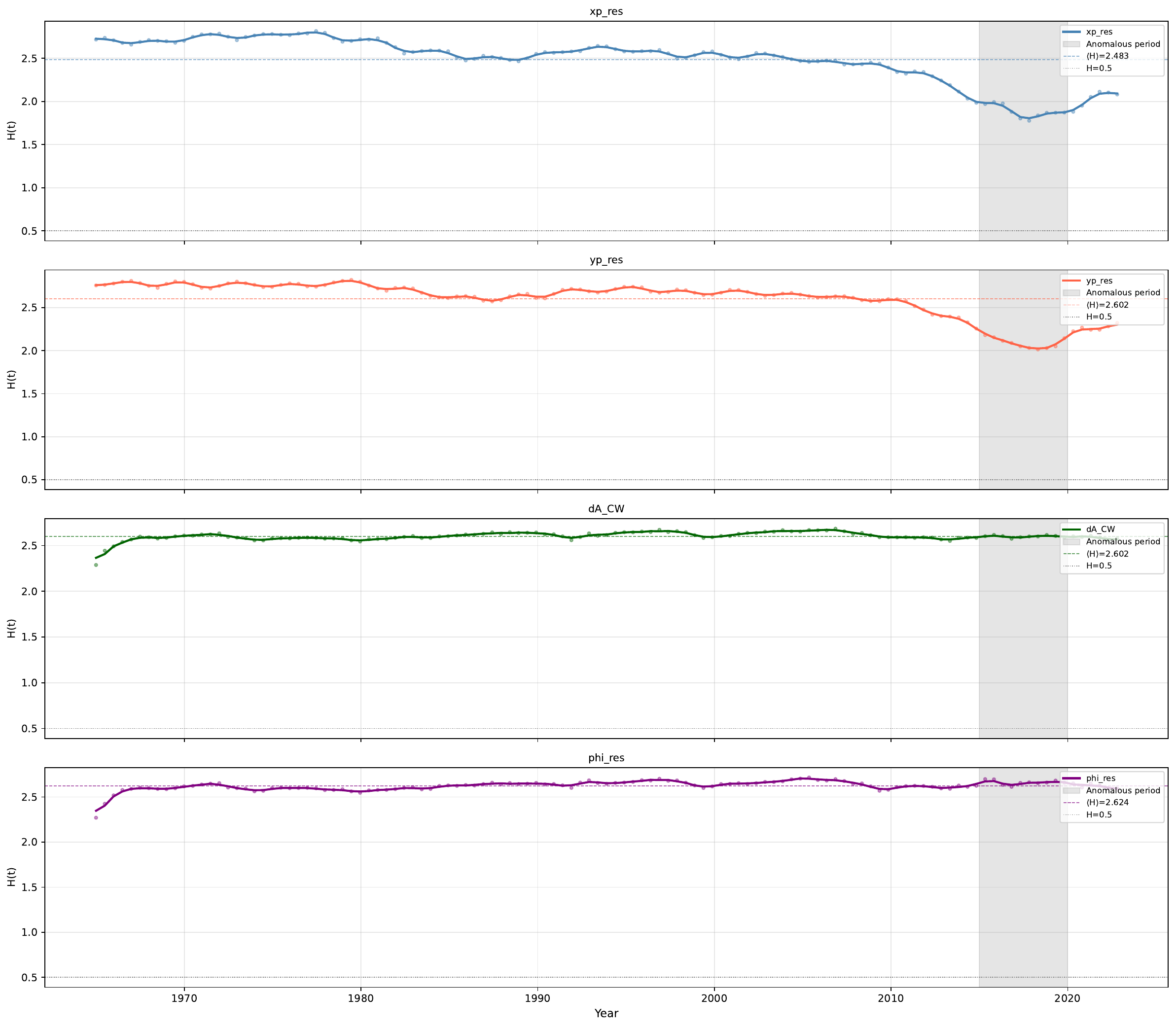}
  \caption{Temporal evolution of the local Hurst exponent
  $H(t)=h(q{=}2,t)$ estimated by sliding-window MFDFA
  (window $W=6$~yr, step $\Delta t=0.5$~yr) for
  $x_p^{\rm res}$, $y_p^{\rm res}$, $\delta A_{\rm CW}$ and
  $\phi_{\rm res}$. The grey band marks the anomalous period
  2015--2020.}
  \label{fig:H_sliding}
\end{figure}

The sliding-window MFDFA reveals strong non-stationarity in the
long-range persistence of the polar motion residuals. The time
evolution of the local Hurst exponent $H(t)=h(q=2,t)$ for the four
series is shown in Fig.~\ref{fig:H_sliding}. For the geometric
residuals $x_p^{\rm res}$ and $y_p^{\rm res}$, $H(t)$ decreases
progressively from values $\sim2.65$--$2.73$ during the pre-active
and active periods (1965--2010) to $\sim1.88$--$2.10$ during the
anomalous period 2015--2020. The decline begins about
5--10~years before the amplitude collapse, around 2005--2010,
suggesting that the loss of long-range persistence precedes the
near-disappearance of the Chandler wobble amplitude.

\begin{table}[htbp]
\centering
\small
\caption{Mean $\pm$ Std of the generalized Hurst exponent $H(t)=h(q=2,t)$
computed by sliding-window MFDFA (window = 6 yr, step = 0.5 yr) for each
temporal segment. Superscripts indicate statistical significance of
differences with respect to the anomalous period (Mann--Whitney $U$ test).
The last column lists the number of windows $n$ in each segment
(cf.\ Table~\ref{tab:segments}).}
\label{tab:H}
\begin{tabular}{lccccr}
\hline\hline
Segment &
$x_p^{\rm res}$ &
$y_p^{\rm res}$ &
$\delta A_{\rm CW}$ &
$\phi_{\rm res}$ \\
\hline
Pre-active (1965--1990) &
$2.669\pm0.105^{***}$ &
$2.725\pm0.071^{***}$ &
$2.589\pm0.037$ &
$2.600\pm0.040^{***}$ &
\\[1ex]
Active (1990--2010) &
$2.529\pm0.064^{***}$ &
$2.664\pm0.044^{***}$ &
$2.632\pm0.029^{**}$ &
$2.656\pm0.033$ &
\\[1ex]
Decline (2010--2015) &
$2.221\pm0.135^{***}$ &
$2.435\pm0.107^{***}$ &
$2.582\pm0.013^{**}$ &
$2.614\pm0.012^{***}$ &
\\[1ex]
Anomalous (2015--2020) &
$1.884\pm0.070$ &
$2.098\pm0.070$ &
$2.600\pm0.012$ &
$2.660\pm0.025$ &
\\[1ex]
Post-anomalous (2020--2024) &
$2.060\pm0.065^{**}$ &
$2.271\pm0.033^{***}$ &
$2.584\pm0.020$ &
$2.613\pm0.024^{**}$ &
\\
\hline
\multicolumn{6}{l}{\footnotesize $^{*}p<0.05$, $^{**}p<0.01$, $^{***}p<0.001$ (Mann--Whitney vs Anomalous)}\\
\hline\hline
\end{tabular}
\end{table}

Segment-wise statistics of $H(t)$ are summarised in
Table~\ref{tab:H}. For $x_p^{\rm res}$ and $y_p^{\rm res}$ the mean
values of $H(t)$ in the anomalous segment are lower by
$\Delta H\simeq0.6$--$0.8$ compared with the pre-active and active
epochs, and the corresponding Mann--Whitney $U$ tests yield highly
significant differences ($p<0.001$ in all cases). In contrast, the
amplitude- and phase-related series $\delta A_{\rm CW}$ and
$\phi_{\rm res}$ do not exhibit statistically robust changes in
$H(t)$ across segments: their mean values remain close to
$H(t)\simeq2.60$ in all epochs, with non-significant or only weakly
significant differences relative to the anomalous period. Taken
together, Fig.~\ref{fig:H_sliding} and Table~\ref{tab:H} indicate
that the persistence collapse is specific to the geometric polar
motion components and does not affect the multiscale structure of
the amplitude and phase fluctuations.

\subsection{Temporal evolution of $\Delta\alpha(t)$}
\label{sec:results_Dalpha}

\begin{figure}[htbp]
  \centering
  \includegraphics[width=\linewidth]{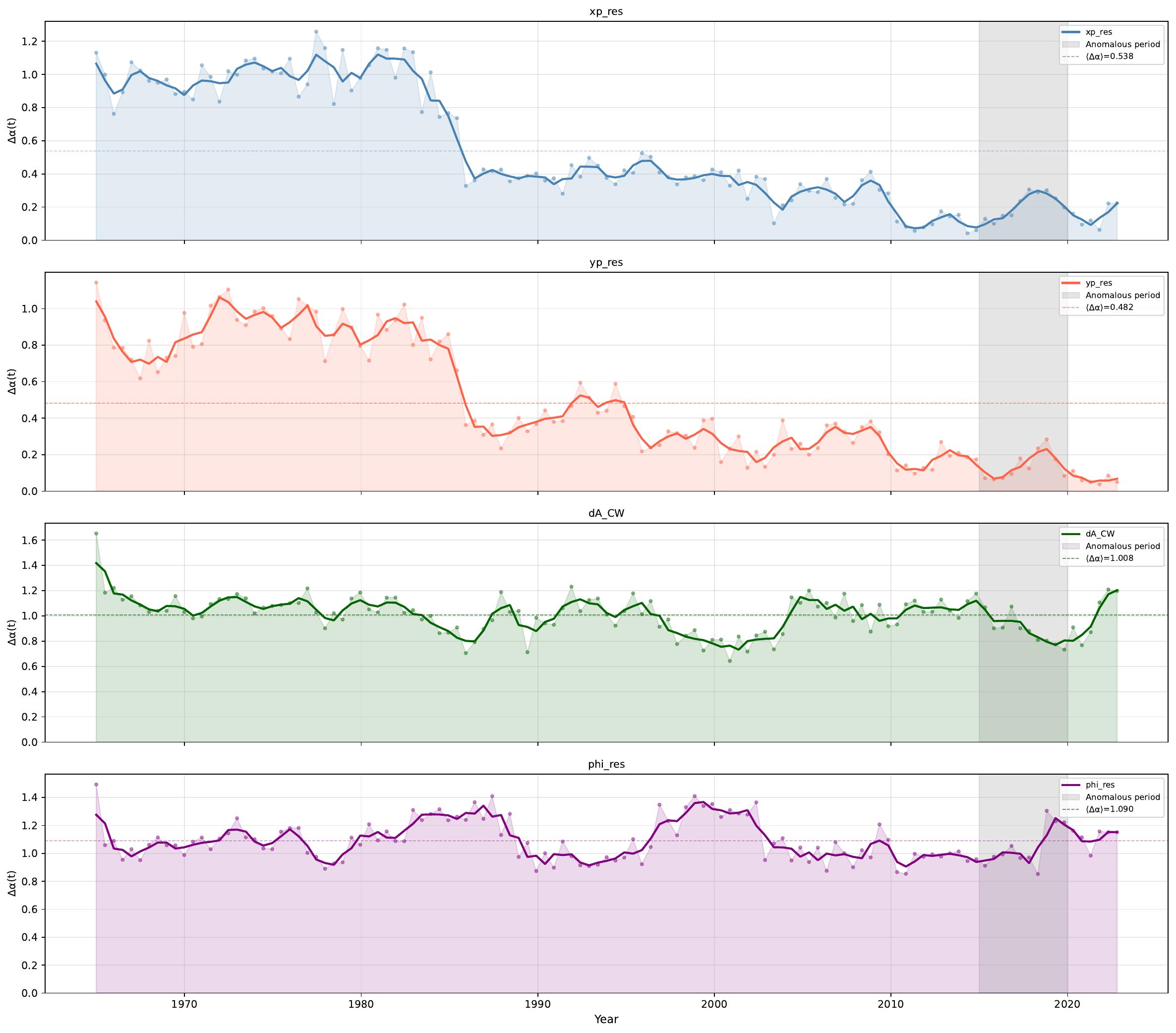}
  \caption{Temporal evolution of the multifractal spectral width
  $\Delta\alpha(t)$ from sliding-window MFDFA for the four analysed
  series. The grey band marks the anomalous period 2015--2020.}
  \label{fig:Da_sliding}
\end{figure}

The multifractal spectral width $\Delta\alpha(t)$ also exhibits
pronounced temporal variability (Fig.~\ref{fig:Da_sliding}). For the
polar motion residuals $x_p^{\rm res}$ and $y_p^{\rm res}$,
$\Delta\alpha(t)$ decreases steadily across the record, from values
$\sim0.8$--$0.9$ during the pre-active period to
$\sim0.1$--$0.2$ during and after the anomalous epoch, indicating
a long-term collapse of multifractal complexity in the geometric
polar motion. In contrast, the amplitude- and phase-related series
$\delta A_{\rm CW}$ and $\phi_{\rm res}$ maintain broad spectral
widths $\Delta\alpha(t)\sim1.0$ throughout, with only modest
fluctuations around their long-term means.

\begin{table}[htbp]
\centering
\small
\caption{Mean $\pm$ Std of the multifractal spectral width
$\Delta\alpha(t) = \alpha_{\max}(t)-\alpha_{\min}(t)$ computed by
sliding-window MFDFA (window = 6 yr, step = 0.5 yr) for each
temporal segment. Superscripts indicate statistical significance of
differences with respect to the anomalous period (Mann--Whitney $U$ test).}
\label{tab:Da}
\begin{tabular}{lccccr}
\hline\hline
Segment &
$x_p^{\rm res}$ &
$y_p^{\rm res}$ &
$\delta A_{\rm CW}$ &
$\phi_{\rm res}$ &
\\
\hline
Pre-active (1965--1990) &
$0.874\pm0.259^{***}$ &
$0.776\pm0.235^{***}$ &
$1.040\pm0.118$ &
$1.113\pm0.124$ \\[1ex]
Active (1990--2010) &
$0.353\pm0.088^{***}$ &
$0.327\pm0.115^{***}$ &
$0.968\pm0.150$ &
$1.089\pm0.162$ \\[1ex]
Decline (2010--2015) &
$0.100\pm0.045^{**}$ &
$0.163\pm0.054^{**}$ &
$1.066\pm0.075$ &
$0.958\pm0.056$ \\[1ex]
Anomalous (2015--2020) &
$0.207\pm0.074$ &
$0.136\pm0.073$ &
$0.887\pm0.109$ &
$1.058\pm0.148$  \\[1ex]
Post-anomalous (2020--2024) &
$0.145\pm0.075^{**}$ &
$0.057\pm0.018^{***}$ &
$1.030\pm0.200$ &
$1.112\pm0.074$ \\
\hline
\multicolumn{6}{l}{\footnotesize $^{*}p<0.05$, $^{**}p<0.01$, $^{***}p<0.001$ (Mann--Whitney vs Anomalous)}\\
\hline\hline
\end{tabular}
\end{table}

Table~\ref{tab:Da} summarises the segment-wise statistics of
$\Delta\alpha(t)$. For $x_p^{\rm res}$ and $y_p^{\rm res}$ the
reductions in $\Delta\alpha(t)$ from the pre-active and active
segments to the anomalous and post-anomalous periods are large and
statistically significant (Mann--Whitney $U$ tests with
$p<0.01$--$0.001$ in most comparisons). By contrast,
$\delta A_{\rm CW}$ and $\phi_{\rm res}$ do not show any systematic
trend in $\Delta\alpha(t)$ across segments; their segmental means
remain within the range $0.9$--$1.1$ and the differences relative to
the anomalous period are not statistically robust. The combined
behaviour of $H(t)$ and $\Delta\alpha(t)$ thus points to a
progressive simplification of the multifractal structure of the
geometric polar motion, while the amplitude and phase variables
retain a remarkably stable multifractal complexity across all
epochs, including the 2015--2020 quiescence.

\begin{figure}[htbp]
  \centering
  \includegraphics[width=\linewidth]{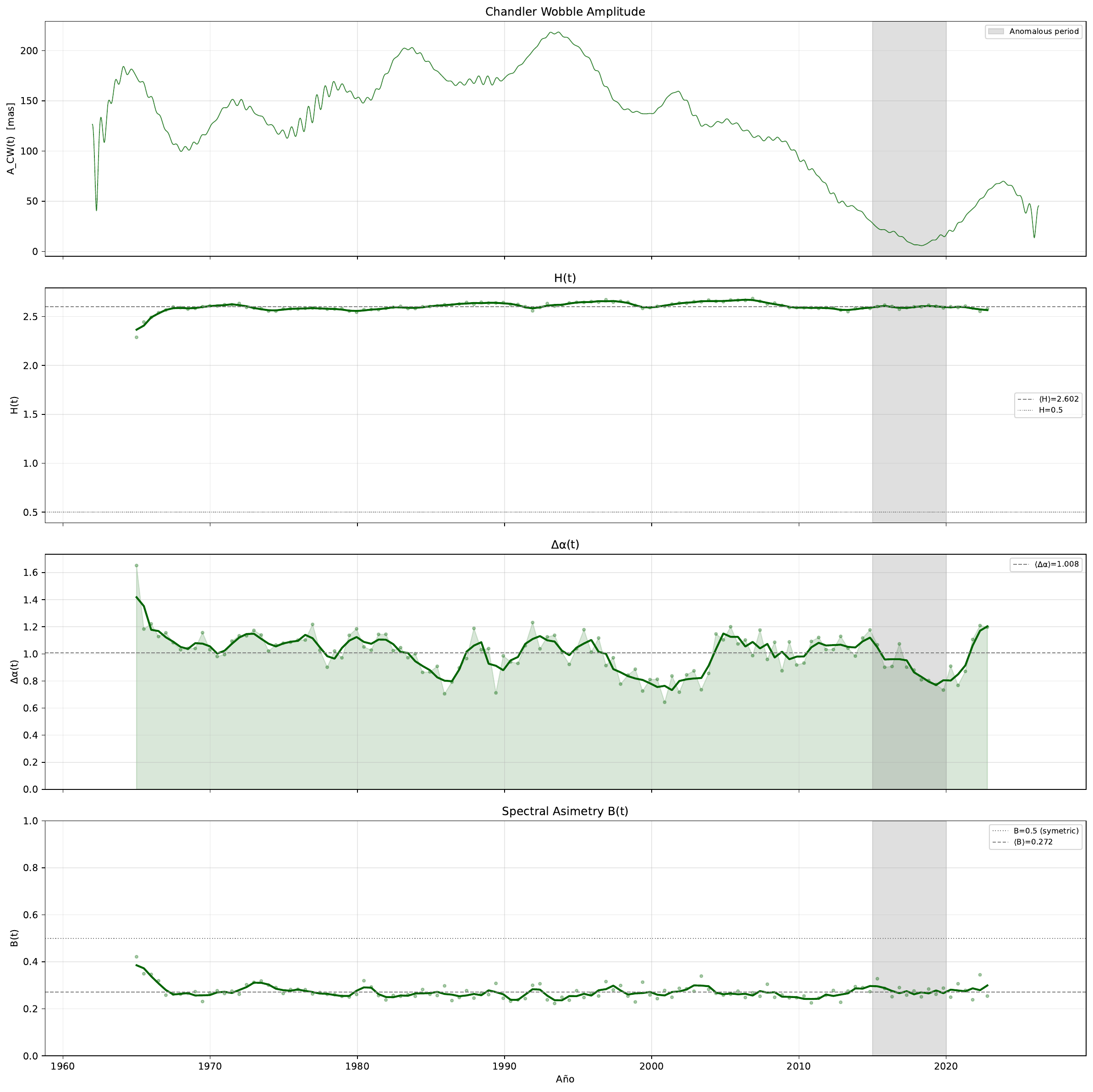}
  \caption{Temporal evolution of the multifractal complexity of the
  Chandler wobble amplitude increments $\delta A_{\rm CW}(t)$.
  Top panel: amplitude envelope $A_{\rm CW}(t)$ (context), showing
  the long-term modulation, the 2015--2020 quiescence and the
  subsequent re-excitation. Second panel: local Hurst exponent
  $H(t)=h(q{=}2,t)$ from sliding-window MFDFA
  (window $W=6$~yr, step $\Delta t=0.5$~yr).
  Third panel: local multifractal spectral width
  $\Delta\alpha(t)$, with the long-term mean highlighted.
  Bottom panel: spectral asymmetry $B(t)$, where values below
  0.5 indicate dominance of large fluctuations in the multifractal
  spectrum. The shaded band marks the anomalous period 2015--2020.}
  \label{fig:clave_dACW}
\end{figure}

For completeness, Fig.~\ref{fig:clave_dACW} shows the full
sliding-window evolution of $H(t)$, $\Delta\alpha(t)$ and $B(t)$
for the amplitude increments $\delta A_{\rm CW}$, together with the
amplitude envelope $A_{\rm CW}(t)$ for context. The plots confirm
that, despite strong modulation of $A_{\rm CW}(t)$, the local
multifractal metrics of $\delta A_{\rm CW}$ remain comparatively
stable across the entire record, including the 2015--2020
quiescence, reinforcing the decoupling between amplitude dynamics
and the geometric polar motion signal.

\subsection{Distributions across segments}
\label{sec:results_boxplots}

\begin{figure}[htbp]
  \centering
  \includegraphics[width=\linewidth]{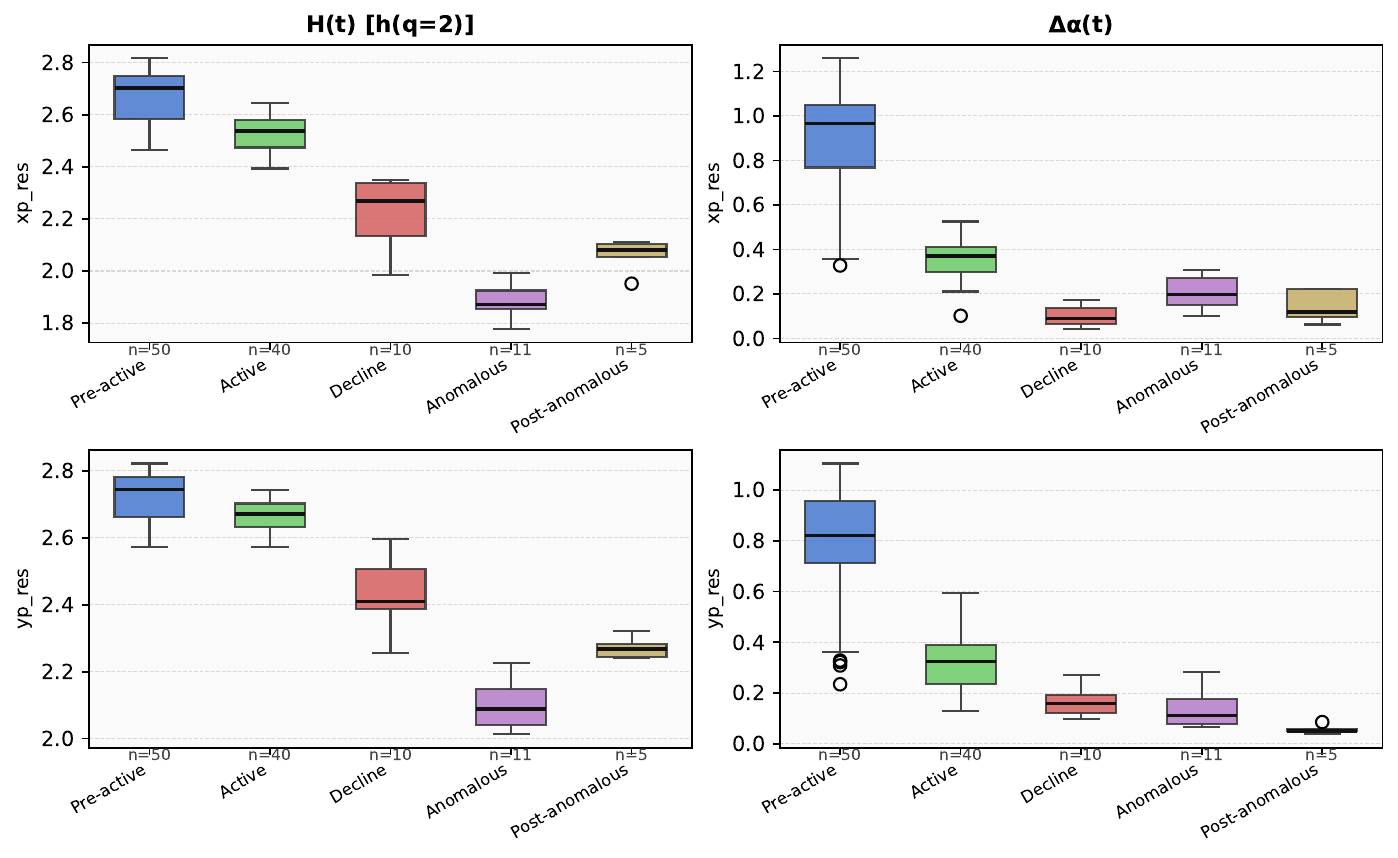}
  \caption{Boxplots of the sliding-window Hurst exponent $H(t)$ (left column)
and multifractal spectral width $\Delta\alpha(t)$ (right column) for
the Chandler wobble residuals $x_p^{\rm res}$ (top row) and
$y_p^{\rm res}$ (bottom row), grouped by the five temporal segments
defined in Table~\ref{tab:segments}. Colours indicate segments
(Pre-active, Active, Decline, Anomalous, Post-anomalous).}
  \label{fig:boxplots_tramos}
\end{figure}

To visualise the full distributions of the sliding-window metrics
across temporal segments, Fig.~\ref{fig:boxplots_tramos} presents
boxplots of $H(t)$ and $\Delta\alpha(t)$ for the Chandler wobble
residuals $x_p^{\rm res}$ and $y_p^{\rm res}$. Each box summarises
the distribution of window-based estimates within a given segment,
with colours matching those used in Fig.~\ref{fig:H_sliding} and
Fig.~\ref{fig:Da_sliding} and the segment definitions in
Table~\ref{tab:segments}. The boxes for the anomalous period
(2015--2020) are markedly shifted towards lower $H(t)$ and
$\Delta\alpha(t)$ for $x_p^{\rm res}$ and $y_p^{\rm res}$, while the
pre-active and active segments exhibit broader, higher-valued
distributions.

The boxplots emphasise two key features already apparent from
Tables~\ref{tab:H} and \ref{tab:Da}. First, the anomalous and
post-anomalous segments show not only lower mean values of $H(t)$
and $\Delta\alpha(t)$ for the geometric residuals, but also a
substantial contraction of the interquartile ranges, indicating a
more homogeneous, less complex fluctuation structure during and
after the 2015--2020 quiescence. Second, the amplitude- and
phase-related variables (not shown in this figure) do not exhibit
comparable shifts: their boxplots remain centred around nearly
constant values of $H(t)$ and $\Delta\alpha(t)$ across all segments,
consistent with the segment-wise statistics in
Tables~\ref{tab:H} and \ref{tab:Da}. Together, these results confirm
that the persistence and multifractal-complexity collapse is
confined to the geometric polar motion components and does not
extend to the multiscale structure of the amplitude and phase
fluctuations.

\begin{figure}[htbp]
  \centering
  \includegraphics[width=\linewidth]{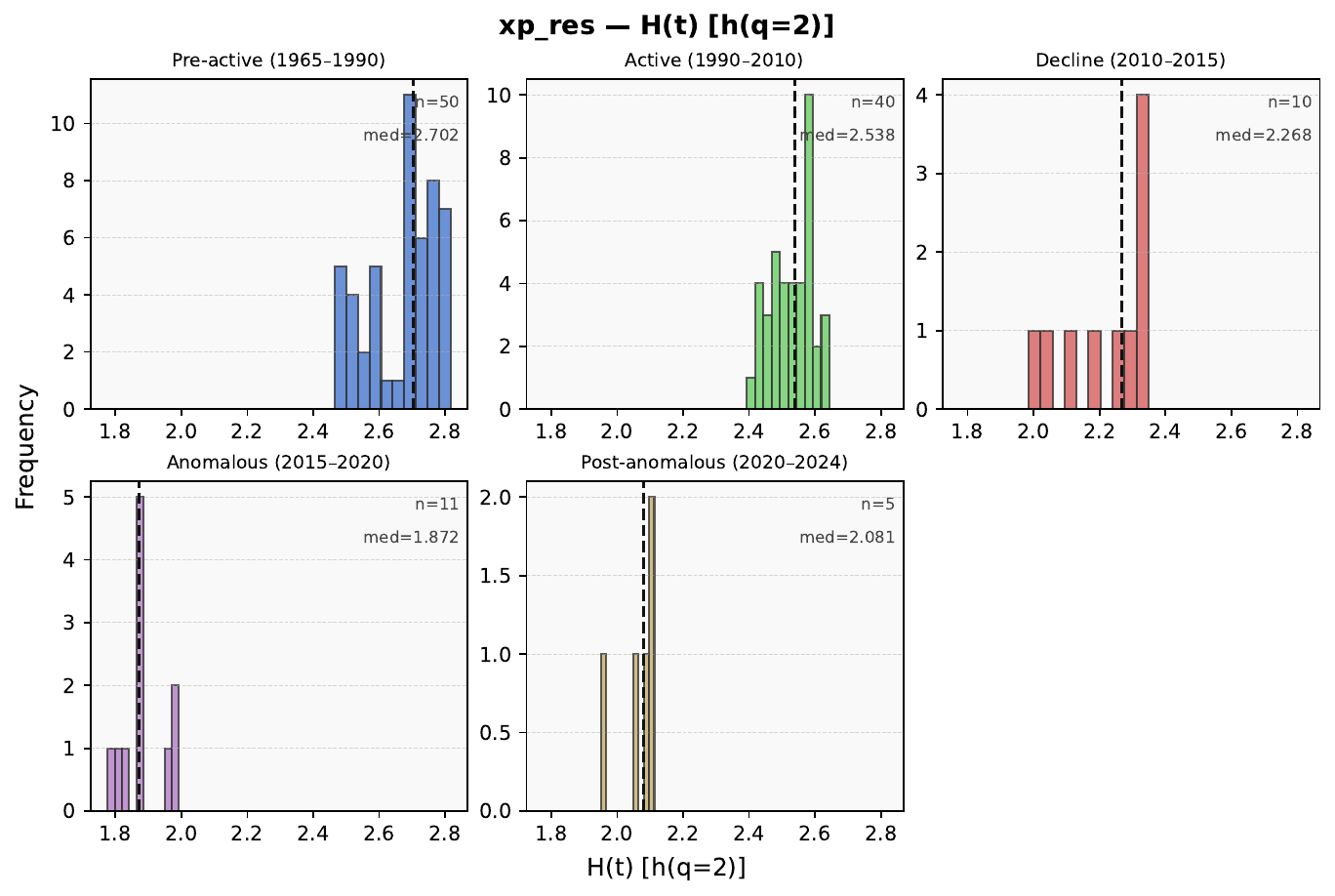} 
  \caption{Segment-wise mean$\pm$Std of (left) $H(t)$,
  (centre) $\Delta\alpha(t)$ and (right) $B(t)$ for all four series.
  The shaded band marks the anomalous period 2015--2020.}
  \label{fig:barras_tramos}
\end{figure}

As a robustness check against potential temporal smearing, we repeated the
analysis by labelling each sliding window by its end time $t_{\mathrm{end}}$,
so that no window ``sees'' the future relative to the segment under
consideration. Under this alternative time parametrisation, the segment-wise
means of $H(t)$ and $\Delta\alpha(t)$ change moderately in magnitude, but the
qualitative pattern remains: during the anomalous segment 2015--2020, the
values of $H(t)$ stay clearly below those of the pre-active and active
segments, while $\Delta\alpha(t)$ remains in the low-to-intermediate range
compared to those same reference periods. The temporal re-labelling neither
shifts the anomaly into the subsequent segment nor removes it, which supports
that the early-warning signal is not a mere artefact of sliding-window
smearing. Under this end‑time parametrisation, the segment‑wise means and Mann–Whitney tests (see Table \ref{tab:xp_H_center_vs_end}) confirm that the 2015–2020 segment remains statistically distinct from both the pre‑active and active segments.

\begin{table}[ht]
  \centering
  \caption{Robustness of segment-wise $H(t)$ estimates for
  $x_p^{\mathrm{res}}$ under two time labellings of the sliding windows:
  by window centre $t_{\mathrm{center}}$ and by window end
  $t_{\mathrm{end}}$. We report the number of windows and the segment-wise
  means of $H(t)$ under each labelling, together with their difference
  $\Delta = H_{\mathrm{end}} - H_{\mathrm{center}}$.}
  \label{tab:xp_H_center_vs_end}
  \begin{tabular}{lrrrrr}
    \hline
    Segment & $n_{\mathrm{center}}$ & $H(t_{\mathrm{center}})$
            & $n_{\mathrm{end}}$    & $H(t_{\mathrm{end}})$ & $\Delta$ \\
    \hline
    Pre-active (1965--1990)     & 50 & 2.6686 & 45 & 2.6918 & +0.0232 \\
    Active (1990--2010)         & 40 & 2.5293 & 40 & 2.5404 & +0.0111 \\
    Decline (2010--2015)        & 10 & 2.2210 & 10 & 2.3924 & +0.1714 \\
    Anomalous (2015--2020)      & 11 & 1.8844 & 11 & 2.0437 & +0.1593 \\
    Post-anomalous (2020--2024) &  5 & 2.0603 &  8 & 1.8887 & -0.1716 \\
    \hline
  \end{tabular}
\end{table}

%% ============================================================
\section{Discussion}
\label{sec:discussion}

\subsection{Persistence collapse and dynamical regime change}

The sliding-window results reveal a robust and systematic decline in
the long-range persistence of the polar motion residuals
$x_p^{\rm res}$ and $y_p^{\rm res}$. The local Hurst estimator
$H(t)\equiv h(q{=}2,t)$ decreases from $\sim$2.65--2.73 in the
pre-active and active periods to $\sim$1.88--2.10 during the
anomalous 2015--2020 interval, a reduction of $\Delta H\approx0.7$
(Figs.~\ref{fig:H_sliding} and \ref{fig:boxplots_tramos};
Table~\ref{tab:H}). We emphasise that these sliding-window
values of $H(t)$ are not directly equivalent to the Hurst exponent
of the original series: because the MFDFA is applied to the
integrated profile and because each 6-yr window contains a
dominant CW component, the local $H(t)$ estimates are systematically
shifted upward relative to the global value $h(2)\approx1.742$
(corresponding to $H_{\rm real}\approx0.742$ for the full 1962--2024
record; see Table~\ref{tab:global_mfdfa}). Accordingly, we base our
interpretation on the \emph{relative} drop $\Delta H\approx0.7$
rather than on the absolute values of $H(t)$.

The associated collapse in multifractal spectral width
($\Delta\alpha$ decreasing from $\sim$0.8--0.9 to $\sim$0.1--0.2;
Figs.~\ref{fig:Da_sliding} and \ref{fig:barras_tramos};
Table~\ref{tab:Da}) indicates that the range of scaling
exponents active in the polar motion dynamics is severely reduced
during and after the CW quiescence. Taken together, the drops in
$H(t)$ and $\Delta\alpha(t)$ are statistically significant relative
to the pre-active and active periods (typically $p<0.01$ using
non-overlapping windows; see Section~\ref{sec:stats_method}).

In physical terms, these findings indicate that the \emph{geometry} of the
Chandler wobble---namely the phase, mean inclination and overall shape of
the trajectory in state space---remains compatible with the historical
behaviour, while the \emph{distribution of variance across time scales} is
radically reorganised. The system continues to precess along a similar mean
orbit, but the redistribution of power between slow and fast components
suppresses the multifractal amplitude and long-range persistence, driving
the dynamics towards a more nearly single-scale, short-memory regime during
and after the quiescent interval.

These results suggest that the 2015--2020 event is not merely an
episode of suppressed CW amplitude but a genuine dynamical regime
change at the level of the scaling structure of polar motion. In a
stochastic-dynamical interpretation, the long-range persistence of
the CW-bearing residuals reflects the presence of coherent, slowly
varying excitation processes (e.g.\ large-scale ocean-bottom pressure
fluctuations) and the system's ability to retain memory of past
forcing over multiple Chandler cycles \cite{Gross2000,Gross2003}.
When the wobble vanishes, this memory is partially erased: the
residual motion becomes dominated by shorter-memory processes whose
correlation structure is closer to that of a near-random walk. One
interpretation invokes the contribution of the CW itself: as a
quasi-periodic coherent oscillation with a damping time of
$\sim$30--70~yr \cite{Gross2000}, the CW introduces strong
correlations at timescales up to several hundred years; when it is
suppressed, these correlations vanish and the scaling structure of
the residuals is reduced accordingly. A complementary interpretation
involves a nonlinear dynamical bifurcation near the quiescence event
\cite{Zotov2023}, in which the system's effective attractor contracts
and its multifractal complexity is reduced.

\subsection{Dual origin of multifractality}

The surrogate-data analysis clarifies the physical origin of the
observed multifractality. The strong reduction of $\Delta\alpha$ in
both shuffled and phase-randomised surrogates
(Fig.~\ref{fig:surrogate_falpha}) and the very large $Z$-scores in
Table~\ref{tab:surrogate_results} demonstrate that multifractality is
not a finite-size artefact but arises from a combination of long-range
correlations and heavy-tailed probability distributions. Shuffling
destroys correlations while preserving the amplitude distribution; the
resulting narrow spectra (small $\Delta\alpha_{\rm sh}$) indicate that
fat tails alone cannot account for the observed complexity. Conversely,
phase randomisation preserves the power spectrum (and thus linear
correlations) while tending to gaussianise the PDF; the intermediate
$\Delta\alpha_{\rm pr}$ values and large $Z_{\rm pr}$ scores show that
LRC alone are also insufficient.

Physically, the LRC component can be traced to the persistence of
ocean-bottom pressure and atmospheric excitation, which are known to
be correlated on multi-year to decadal timescales
\cite{Gross2000,Gross2003}. The fat-tail component likely reflects the
intermittent nature of strong excitation events, such as major ENSO
episodes, abrupt hydrological anomalies, or rapid ice-mass changes
\cite{Adhikari2016,Shi2025,Jeon2025}. The exceptionally large $Z$-scores
for the amplitude envelope $A_{\rm CW}$ ($Z_{\rm sh}=76.6$,
$Z_{\rm pr}=30.0$) underscore the idea that the CW amplitude behaves
as an integrator of multi-scale excitation, accumulating the effects
of both persistent and intermittent forcing into a highly hierarchical,
multifractal signal.

\subsection{Decoupling of geometry and amplitude/phase complexity}

An important outcome of this analysis is the apparent decoupling
between the multifractal behaviour of the geometric polar motion
($x_p^{\rm res}$, $y_p^{\rm res}$) and that of the amplitude and phase
variables ($\delta A_{\rm CW}$, $\phi_{\rm res}$).
While both $H(t)$ and $\Delta\alpha(t)$ exhibit strong, statistically significant
reductions for the residual components during the anomalous period
(Figs.~\ref{fig:boxplots_tramos} and \ref{fig:barras_tramos};
Tables~\ref{tab:H} and \ref{tab:Da}), neither
$\delta A_{\rm CW}$ nor $\phi_{\rm res}$ shows comparable changes.
The amplitude increment series maintains
$\Delta\alpha \approx 1.0$ across all segments, and its Hurst exponent
remains close to $H_{\rm real} \approx 0.6$--0.7; similarly, the phase
residual preserves a broad multifractal spectrum with stable width and
asymmetry.

In physical terms, this decoupling indicates that the \emph{geometry} of the
Chandler wobble---namely the phase, mean inclination, and overall shape of
the trajectory in state space---remains essentially unchanged across
the transition, while the \emph{spectral distribution of variance across
time scales} is reorganised abruptly. In other words, the pole continues
to precess along a similar mean orbit, but the redistribution of power
between slow and fast components suppresses the multifractal amplitude,
driving the system towards a more nearly single-scale regime during
the anomalous episode.

\subsection{Spectral asymmetry and intermittency}

The spectral asymmetry index $B(t)$ remains below 0.5 for all series
and epochs (Fig.~\ref{fig:boxplots_tramos}), indicating that large
fluctuations (negative $q$) consistently dominate the multifractal
spectra. This is characteristic of intermittent dynamics in which
quiescent periods are punctuated by bursts of high activity. For the
polar motion residuals, this can be interpreted as a signature of
occasional strong excitation events superimposed on a background of
weaker, more regular forcing. The absence of a clear long-term trend
in $B(t)$ suggests that the \emph{type} of intermittency remains
similar across epochs; what changes is the overall degree of
persistence and the breadth of the multifractal spectrum.

For the amplitude series, the persistent left-skewed spectra
($B\lesssim0.3$) indicate that extreme amplitude changes play a
disproportionate role in shaping the scaling properties of
$\delta A_{\rm CW}$. This behaviour is consistent with the large
excursions associated with major CW excitation or de-excitation
episodes, such as those preceding and following the 2015--2020
quiescence.

\subsection{Multifractal metrics as potential early-warning indicators}

One of the most intriguing results is that both $H(t)$ and
$\Delta\alpha(t)$ for the polar motion residuals begin to decline
5--10~years before the CW amplitude collapses (Figs.~\ref{fig:H_sliding}
and \ref{fig:Da_sliding}). This temporal ordering suggests that
multifractal metrics may act as early-warning indicators of impending
changes in CW amplitude, analogous to critical-slowing-down signatures
in complex systems approaching bifurcations or tipping points.

At present, this interpretation must remain tentative, as the
instrumental record contains only one clearly documented quiescence
event of this magnitude (the 2015--2020 episode). A decisive test
would require reprocessing historical polar motion series to
reconstruct the earlier low-amplitude episode of the 1920s--1930s and
applying the same MFDFA and surrogate pipeline. Nevertheless, the
present results motivate systematic monitoring of multifractal metrics
as part of operational Earth rotation services, particularly given
their sensitivity to changes in the long-range correlation structure
of the excitation processes.

\subsection{Limitations and future work}

Several limitations of this study suggest avenues for further
investigation. First, the choice of MFDFA parameters (polynomial order
$m=2$, scale range, $q$-range) follows standard practice but is not
unique; although extensive tests show that the main conclusions are
robust, a systematic sensitivity analysis could quantify the
associated uncertainty more rigorously. Second, the surrogate tests
consider only two classes (RS and IAAFT); additional surrogates that
selectively preserve non-linear correlations could help to further
separate linear and nonlinear contributions to multifractality.

Third, the present analysis focuses on univariate series. A natural
extension is to apply multifractal cross-correlation analysis (MF-DCCA)
between $x_p^{\rm res}$, $y_p^{\rm res}$ and external geophysical
drivers such as GRACE-based hydrological mass anomalies or
reanalysis-based ocean-bottom pressure fields. This would permit a
more direct attribution of the multifractal structure of polar motion
to specific forcing mechanisms. Finally, applying the same methodology
to other Earth orientation parameters (e.g.\ LOD, nutation) could
reveal whether similar complexity collapses accompany other aspects of
Earth rotation during the 2010s, thereby providing a more unified
picture of the dynamical response of the Earth system to contemporary
climate-related mass redistribution.

%% ============================================================
\section{Conclusions}
\label{sec:conclusions}

This study presents the first multifractal characterisation of the Chandler wobble using MFDFA applied to 62~years of daily IERS EOP C04
data. In conclusion, the global MFDFA demonstrates that the polar motion residuals and the Chandler wobble amplitude behave as genuine multifractal processes, exhibiting strong complexity ($\Delta\alpha \approx 0.72-1.61$) that is rigorously confirmed by surrogate testing (Z-scores 5-77). Our analysis reveals that this multifractality originates from a dual contribution of long-range temporal correlations—likely tied to correlated ocean-bottom pressure dynamics—and heavy-tailed probability distributions driven by intermittent, sporadic geophysical excitations. Crucially, the anomalous 2015-2020 quiescence is not merely an amplitude suppression but constitutes a qualitative dynamical regime change, marked by a statistically significant collapse in both long-range persistence ($\Delta h(q=2) \approx 0.6-0.8$, $p<0.001$) and multifractal spectral width. Despite this profound structural simplification in the geometric polar motion, the amplitude variability and phase residuals maintain a strikingly constant multifractal complexity throughout all periods, revealing a deep dynamical decoupling between the geometric and amplitude dynamics of the wobble. Finally, we observe that the decline in long-range persistence precedes the amplitude collapse by roughly 5 to 10 years, suggesting that multifractal scaling metrics could serve as powerful early-warning indicators for impending major transitions in Earth's rotational dynamics.

%% ============================================================

\section*{Declaration of Competing Interest}
The authors declare no competing interests.

\section*{Data Availability}
IERSEOP C04 data are freely available at
\url{https://hpiers.obspm.fr/iers/eop/eopc04}. The multifractal
analysis was carried out with the \texttt{MF-toolkit} Python library
\cite{Mendez2026_MFtoolkit}, and the corresponding analysis scripts
built on this library are available from the authors upon reasonable
request.

%% ============================================================
\bibliographystyle{elsarticle-num}
\bibliography{references}

@article{Aoyama2001,
   author = {Yuichi Aoyama and Isao Naito},
   doi = {10.1029/2000JB900460},
   issn = {0148-0227},
   issue = {B5},
   journal = {Journal of Geophysical Research: Solid Earth},
   month = {5},
   pages = {8941-8954},
   title = {Atmospheric excitation of the Chandler wobble, 1983–1998},
   volume = {106},
   year = {2001}
}

@article{Telesca2004,
   author = {Luciano Telesca and Vincenzo Lapenna and Maria Macchiato},
   doi = {10.1016/S0960-0779(03)00188-7},
   issn = {09600779},
   issue = {1},
   journal = {Chaos, Solitons and Fractals},
   pages = {1-15},
   publisher = {Elsevier Ltd},
   title = {Mono- and multi-fractal investigation of scaling properties in temporal patterns of seismic sequences},
   volume = {19},
   year = {2004}
}

@article{Peng1994,
   author = {C.-K. Peng and S. V. Buldyrev and S. Havlin and M. Simons and H. E. Stanley and A. L. Goldberger},
   doi = {10.1103/PhysRevE.49.1685},
   issn = {1063-651X},
   issue = {2},
   journal = {Physical Review E},
   month = {2},
   pages = {1685-1689},
   title = {Mosaic organization of DNA nucleotides},
   volume = {49},
   year = {1994}
}

@article{Adhikari2016,
   author = {Surendra Adhikari and Erik R. Ivins},
   doi = {10.1126/sciadv.1501693},
   issn = {2375-2548},
   issue = {4},
   journal = {Science Advances},
   month = {4},
   title = {Climate-driven polar motion: 2003–2015},
   volume = {2},
   year = {2016}
}

@article{Jeon2025,
   author = {Taehwan Jeon and Ki Weon Seo and Kookhyoun Youm and Jooyoung Eom and Daeha Lee and Jianli Chen and Clark R. Wilson},
   doi = {10.1029/2025GL116191},
   issn = {19448007},
   issue = {18},
   journal = {Geophysical Research Letters},
   month = {9},
   publisher = {John Wiley and Sons Inc},
   title = {Diminished Chandler Wobble After 2015: Link to Mass Anomalies in 2011},
   volume = {52},
   year = {2025}
}

@article{Shi2025,
   author = {Qiqi Shi and Yonghong Zhou and Jianli Chen and Xueqing Xu},
   doi = {10.1007/s00190-025-02021-w},
   issn = {14321394},
   issue = {12},
   journal = {Journal of Geodesy},
   month = {12},
   publisher = {Springer Science and Business Media Deutschland GmbH},
   title = {Recent disappearing and re-excited Earth’s Chandler wobble: contributions from GRACE/GFO hydrological and cryospheric mass changes},
   volume = {99},
   year = {2025}
}

@inproceedings{Zotov2023,
   author = {L. Zotov and Ch. Bizouard and N. Sidorenkov},
   doi = {10.1142/9789811275449_0052},
   month = {5},
   pages = {153-155},
   publisher = {World Scientific Pub Co Pte Ltd},
   title = {CHANDLER WOBBLE AND LOD ANOMALIES IN 2010-2020S},
   year = {2023}
}

@inproceedings{Zotov2024,
  title = {Chandler wobble changes in 2020s},
  author = {Zotov, Leonid and Sidorenkov, Nikolay and Bizouard, Christian},
  booktitle = {15th International Conference and School ``PROBLEMS OF GEOCOSMOS''},
  year = {2024},
  month = {April 22-26},
  address = {St. Petersburg},
  note = {ID: GC2024-SG010}
}

@article{Gross2003,
   author = {Richard S. Gross and Ichiro Fukumori and Dimitris Menemenlis},
   doi = {10.1029/2002jb002143},
   issn = {0148-0227},
   issue = {B8},
   journal = {Journal of Geophysical Research: Solid Earth},
   month = {8},
   publisher = {American Geophysical Union (AGU)},
   title = {Atmospheric and oceanic excitation of the Earth's wobbles during 1980–2000},
   volume = {108},
   year = {2003}
}

@article{Gross2000,
   author = {R. S. Gross},
   doi = {10.1029/2000GL011450},
   issn = {00948276},
   issue = {15},
   journal = {Geophysical Research Letters},
   month = {8},
   pages = {2329-2332},
   publisher = {American Geophysical Union},
   title = {The excitation of the Chandler wobble},
   volume = {27},
   year = {2000}
}

@techReport{Kantelhardt2001,
   author = {Jan W Kantelhardt and Eva Koscielny-Bunde and H A Rego and Shlomo Havlin and Armin Bunde},
   journal = {Physica A},
   pages = {441-454},
   title = {Detecting long-range correlations with detrended ductuation analysis},
   volume = {295},
   url = {www.elsevier.com/locate/physa},
   year = {2001}
}

@techReport{Schreiber2000,
   author = {Thomas Schreiber and Andreas Schmitz},
   journal = {Physica D},
   title = {Surrogate time series},
   volume = {142},
   year = {2000}
}

@techReport{Kantelhardt2002,
   author = {Jan W Kantelhardt and Stephan A Zschiegner and Eva Koscielny-Bunde and Shlomo Havlin and Armin Bunde and H Eugene Stanley},
   journal = {Physica A},
   pages = {87-114},
   title = {Multifractal detrended ductuation analysis of nonstationary time series},
   volume = {316},
   url = {www.elsevier.com/locate/physa},
   year = {2002}
}

@misc{Mendez2026_MFtoolkit,
      title={MF-toolkit: A High-Performance Python Library for Multifractal Analysis with Automated Crossover Detection, Source Identification and Application to Gravitational Waves Data}, 
      author={Nahuel Mendez and Maria Cristina Mariani Maria Pia Beccar-Varela and Osei Tweneboah and Sebastian Jaroszewicz},
      year={2026},
      eprint={2604.16257},
      archivePrefix={arXiv},
      primaryClass={cond-mat.stat-mech},
      url={https://arxiv.org/abs/2604.16257}, 
}

@misc{Jaroszewicz2026_IsingFSS,
      title={Resolving Spurious Multifractality in Discrete Systems: A Finite-Size Scaling Protocol for MFDFA in the 2D Ising Model}, 
      author={Sebastian Jaroszewicz and Nahuel Mendez and Maria P. Beccar-Varela and Maria Cristina Mariani},
      year={2026},
      eprint={2603.04609},
      archivePrefix={arXiv},
      primaryClass={cond-mat.stat-mech},
      url={https://arxiv.org/abs/2603.04609}, 
}

@article{Mendez2026_humpback,
author = {Nahuel Mendez and Sebastián Jaroszewicz and Osei K. Tweneboah and Maria P. Beccar-Varela and Maria Cristina Mariani},
title = {Characterising the interplay of dynamics and artefacts: a multifractal analysis of historical humpback whale recordings},
journal = {Bioacoustics},
volume = {35},
number = {2},
pages = {220--239},
year = {2026},
publisher = {Taylor \& Francis},
doi = {10.1080/09524622.2026.2629322},
URL = { 
         https://doi.org/10.1080/09524622.2026.2629322
},
eprint = { 
            https://doi.org/10.1080/09524622.2026.2629322
}
}

@incollection{Gross2015,
  author       = {R. S. Gross},
  title        = {Earth Rotation Variations --- Long Period},
  editor       = {G. Schubert},
  booktitle    = {Treatise on Geophysics},
  edition      = {2},
  volume       = {3},
  series       = {Geodesy},
  pages        = {215--261},
  publisher    = {Elsevier},
  address      = {Oxford},
  year         = {2015},
  doi          = {10.1016/B978-0-444-53802-4.00017-1}
}

@book{Lambeck, place={Cambridge}, series={Cambridge Monographs on Mechanics}, title={The Earth’s Variable Rotation: Geophysical Causes and Consequences}, publisher={Cambridge University Press}, author={Lambeck, Kurt}, year={1980}, collection={Cambridge Monographs on Mechanics}}

\end{document}